\documentclass[aps,twocolumn,showpacs,floatfix]{revtex4}
\usepackage{amssymb,amsbsy,psfig,graphicx}

\def\be{\begin{equation}}
\def\ee{\end{equation}}
\def\bea{\begin{eqnarray}}
\def\eea{\end{eqnarray}}
\begin{document}

%%%%%%%%%%%%%%%%%%%%%%%%%%

\title{Structure of multiphoton quantum optics. II. Bipartite systems, \\
physical processes, and heterodyne squeezed states}

%%%%%%%%%%%%%%%%%%%%%%%%%%%

\author{Fabio Dell'Anno}\email{dellanno@sa.infn.it}
\author{Silvio De Siena}\email{desiena@sa.infn.it}
\author{Fabrizio Illuminati}\email{illuminati@sa.infn.it}

\vspace{0.2cm}

\affiliation{Dipartimento di Fisica ``E. R. Caianiello'',
Universit\`a di Salerno, INFM UdR Salerno, INFN Sez. Napoli, G. C.
Salerno, Via S. Allende, 84081 Baronissi (SA), Italy}

%%%%%%%%%%%%%%%%%%%%%%%%%%%


\date{August 5, 2003}

\begin{abstract}
Extending the scheme developed for a single
mode of the electromagnetic field in the preceding paper
{\bf ``Structure of multiphoton quantum optics. I. Canonical formalism
and homodyne squeezed states''}, we introduce two-mode
nonlinear canonical transformations depending on two heterodyne
mixing angles. They are defined in terms of
hermitian nonlinear functions that realize heterodyne superpositions of
conjugate quadratures of bipartite systems.
The canonical transformations diagonalize a class
of Hamiltonians describing non degenerate and degenerate
multiphoton processes. We determine the coherent states associated to
the canonical transformations, which generalize the non degenerate two--photon
squeezed states. Such heterodyne multiphoton squeezed
are defined as the simultaneous
eigenstates of the transformed, coupled annihilation operators. They are
generated by nonlinear unitary evolutions acting on two-mode squeezed states.
They are non Gaussian, highly non classical, entangled states.
For a quadratic nonlinearity the
heterodyne multiphoton squeezed states define two--mode
cubic phase states. The statistical properties of
these states can be widely adjusted by tuning the heterodyne
mixing angles, the phases of
the nonlinear couplings, as well as the strength of the
nonlinearity. For quadratic nonlinearity, we study the higher-order
contributions to the susceptibility in nonlinear media and we suggest
possible experimental realizations of multiphoton conversion processes
generating the cubic-phase heterodyne squeezed states.
\end{abstract}

\maketitle

\section{Introduction}

In this paper we extend to bipartite systems of pairs of
correlated modes the multiphoton canonical
formalism developed for a single mode of the electromagnetic field in
the companion paper {\bf "Structure of multiphoton
quantum optics. I. Canonical formalism and homodyne squeezed
states"} \cite{paper1}, which, from now on, we will refer to as
Part I. Preliminary to the description of the multiphoton
canonical formalism we need to discuss briefly
the well established formalism of two-mode quantum optics
\cite{yuen,caves}. Two--mode non degenerate
squeezed states are
generated by second order susceptibility contributions excited by
laser shots on nonlinear optical media. Due to the second order
nonlinearity, the frequency $\Omega$ of a high energy laser splits
inside the crystal into a pair of frequencies $\omega_1, \omega_2$
($ \Omega = \omega_1+ \omega_2$) associated to correlated modes of
the electromagnetic field (non degenerate down conversion
process), with the reversed up conversion process of recomposition of the two
frequencies being allowed as well. The degenerate limit $\omega_1 =
\omega_2$ can be considered, with the splitted
frequencies being associated to a single mode of the electromagnetic
field. Two--photon squeezed states are the coherent states of the
Hamiltonian which describes the simultaneous creation or
annihilation of a photon in each of the correlated modes (or of
two photon in a single mode for the degenerate case). The
two--photon Hamiltonian is diagonalized by coupled linear
canonical transformations. The two--photon squeezed states can
be obtained as common eigenstates of the two--mode transformed
variables. They can be generated as well by applying a (squeezing)
unitary operator on the two-mode vacuum or on the two-mode
coherent state; the unitary operator moves the original mode
operators to the canonically transformed variables.

When considering possible extensions of the above scheme to
multiphoton processes, one natural choice is to consider
down--conversion processes in which the carrying laser frequency
splits into $n$ frequencies, giving rise to the creation (and
later annihilation) of $n$ photons in $n$ correlated modes of the
field (or in a single mode of the field in the degenerate limit).
However, the related Hamiltonians cannot be exactly diagonalized,
and an exact canonical scheme cannot be implemented. In this
paper we follow a different
route, outlined in Part I for single-mode systems.
We consider classes of multiphoton processes
in multimode systems such that the number of photons involved in
the processes does not correspond in general to the number of excited modes.
In such instances, the fully degenerate limit entails that
the interaction terms in the Hamiltonians do not in general reduce
to simple powers of the creation and
annihilation operators. In Part I \cite{paper1}, we have
introduced degenerate, canonical $n$-photon generalizations of the
degenerate, canonical two--photon processes. The single-mode multiphoton
generalization has been obtained by using nonlinear operator
functions of {\it homodyne} superpositions of conjugate
quadratures.

To the end of generalizing
the canonical formalism to two-mode, non degenerate multiphoton processes,
we consider nonlinear operator functions of {\it
heterodyne} superpositions of the two--mode variables, obtaining
coupled two--mode transformed operators with a nonlinear
dependence on the original operators. As in the one--mode
instance, we show that the canonical conditions reduce to simple
algebraic constraints only on the complex coefficients of the
transformations, while they are independent of
the form of the nonlinear functions.

Upon introducing multiphoton canonical transformations for bipartite
systems, we analyze the problems regarding the
multiphoton Hamiltonians induced by the transformations, their
physical interest, and their experimental realizability.
We determine the multiphoton coherent states associated to
the transformations, and study their statistical properties. Finally,
we show that the two-mode multiphoton coherent states can be generated
by acting with a unitary operator on
the two--mode vacuum. The structure of the two-mode multiphoton canonical
transformations does not allow the realization of the associated
coherent states by the displacement operator method, i.e. by
the unitary evolution with the interaction Hamiltonian. This problem
is similar to that of mapping interacting fields onto free fields in
Quantum Field Theory for nonlinear systems. We thus consider the coherent states
defined as the simultaneous eigenstates of the coupled canonical
transformations. Exploiting the entangled state
representation we can then compute their eigenfunctions.
The resulting states share important features of the standard coherent states,
such as forming an overcomplete basis in Hilbert space.
These multiphoton coherent states associated to the
nonlinear transformations share squeezing properties as well, and
we thus name them (two--mode) heterodyne multiphoton squeezed states
(HEMPSS). They are non Gaussian, highly non classical, entangled states.
A large part of the present paper is devoted to the characterization and
quantification of their physical properties.

The independence of the canonical transformations
by the form of the nonlinear functions allows in principle to
introduce an infinite number of diagonalizable
Hamiltonians. In analogy to the degenerate case we have
considered the subclass of Hamiltonians involving only algebraic
powers of the heterodyne variables.
This choice defines Hamiltonians associated with
degenerate and degenerate multiphoton processes in two modes
of the electromagnetic field. The higher-order
multiphoton processes are related to higher-order contributions
to the susceptibility in nonlinear media. We then study in detail
only the simplest case of quadratic nonlinearity, which leads to
Hamiltonians describing up to four-photon processes.
We analyze the photon statistics
of the HEMPSS, showing the strong dependence of the
interference in phase space by the competing phases entering in
the nonlinearity, and by the heterodyne angles. We
suggest schemes for a possible experimental
realization of the HEMPSS by exploiting contributions
to the susceptibility up to the fifth order in
nonlinear media. We determine the
unitary operator acting on the vacuum to generate
the HEMPSS: its form suggests alternative routes to
their possible experimental realization.

We remark that multimode nonlinear canonical transformations
of the form introduced in the present paper can be defined
in principle for generic bosonic systems of atomic,
molecular, and condensed matter physics.
In such cases however the constraints on
the standard form of the kinetic energy term and on
the conservation of the total number of particles play
a crucial role and must be taken carefully into account.
A thorough analysis of the relevance of the nonlinear canonical formalisms
for material systems lies thus outside the scope of the present paper.

The paper is organized as follows. In Section~\ref{nlcntr} we
introduce the two--mode nonlinear canonical transformations and
the associated multiphoton Hamiltonians.
In Sections~\ref{entstrep} and \ref{eigenveq} we review the
entangled state representation, and exploit it to determine
the coherent states of the transformations, defined
as the eigenstates of the transformed annihilation operators
(heterodyne multiphoton squeezed states, or HEMPSS). In Section~\ref{Uop} we
derive the form of the unitary operators generating the HEMPSS
by acting on the two--mode vacuum. In Section~\ref{stat}
we study the photon number distribution of the
two--mode multiphoton squeezed states. In Section~\ref{nlqopt} we
review the theory of quantized fields in nonlinear media, and in
Section~\ref{expsetup} we present a proposal for a possible experimental
generation of the HEMPSS.
Finally, in Section~\ref{sumout} we draw our conclusions and
discuss further future developments.

\section{Two--mode multiphoton canonical transformations via
heterodyne photocurrent variables} \label{nlcntr}

In this Section we introduce the two-mode multiphoton canonical
formalism. We consider two (correlated) modes of the quantized
electromagnetic field. The modes are characterized by the
annihilation and creation operators $a_{1}$, $a_{2}$,
$a_{1}^{\dag}$ and $a_{2}^{\dag}$, obeying the canonical
commutation relations
\be
[a_{i},a_{j}^{\dag}]= \delta_{ij} \; ,
\quad \quad [a_{i},a_{j}]=0 \; .
\label{fvcr}
\ee
We begin by recalling that the form of the two-mode, linear
canonical squeezing transformations is \cite{caves}:
\bea
b_{1} & = & \mu a_{1} + \nu a_{2}^{\dagger} \, ,\nonumber \\
\label{sst} \\
b_{2} & = & \mu a_{2} + \nu a_{1}^{\dagger} \, , \nonumber
\eea
with $\mu$, $\nu$ complex numbers.
As is well known, the above transformations are canonical if
\begin{equation}
|\mu|^{2} - |\nu|^{2} = 1 \; ,
\label{lcc}
\end{equation}
and the constraint is automatically satisfied by the
parametrization
\be
\mu = \cosh r \; , \quad     \nu = \sinh r e^{i
\phi} \; ,
\label{lp}
\ee
where $r$ is the squeezing parameter.
The transformations Eqs.~(\ref{sst}) can be obtained from the
original mode operators $a_{1}$, $a_{2}$ by acting with the
unitary operator \cite{caves}
\be
S_{12}(\zeta) = \exp(\zeta a_{2}^{\dagger} a_{1}^{\dagger} -
{\zeta}^{*} a_{2} a_{1}) \, , \, \zeta = r e^{i\phi} \, .
\label{luo}
\ee
The transformed variables $b_{1}, b_{2}$ define the diagonalized
Hamiltonian
\be
H=b_{1}^{\dag}b_{1}+b_{2}^{\dag}b_{2} \, ,
\label{gmH}
\ee
which can be written in terms of the fundamental mode
operators as:
\bea
H & = & (|\mu|^{2}+|\nu|^{2})(a_{1}^{\dag}a_{1} + a_{2}^{\dag}a_{2})
+ 2|\nu|^{2} \nonumber \\
& + & 2\mu^{*} \nu a_{1}^{\dag}a_{2}^{\dag} +2\mu \nu^{*} a_{1} a_{2}
\, .
\label{lsH}
\eea
The scheme defined by relations Eqs.~(\ref{sst}), (\ref{lp}),
(\ref{luo}), (\ref{lsH}) describes
non degenerate two-photon down-conversion processes.

We wish to extend the linear canonical scheme
to provide an Hamiltonian description of multiphoton
processes. Let us define a new set of
transformed variables $b_{1}$ and $b_{2}$ by
\bea
b_{1} & = & \mu a_{1} + \nu a_{2}^{\dagger} + \gamma
F(a_{1}^{\dagger},a_{2}) \; ,\nonumber \\
\label{nlct} \\
b_{2} & = & \mu a_{2} + \nu a_{1}^{\dagger} + \chi
F^{\dagger}(a_{1}^{\dagger},a_{2}) \; . \nonumber
\eea
Here $\mu$, $\nu$, $\gamma$ and $\chi$ are complex numbers.
The generalization of the two--photon linear canonical transformations
is obtained by introducing a \emph{nonlinear}, operator--valued
function $F$ of the creation and annihilation
operators. The function $F$ must satisfy some regularity conditions.
Even so, the canonical conditions may not in general be satisfied by
a completely arbitrary, though well behaved, function.
We will now introduce two crucial characterizations of $F$,
allowing to select a large class of functions that
do satisfy the canonical constraints.
First, we assume the argument of $F$ to be of the form
\be
\alpha a_{1}^{\dag}+\beta^{*} a_{2} \; \; , \; \; \alpha , \beta \in
{\bf C} \, .
\label{cfF1}
\ee
We next require that
\be
F^{\dag}(Z) = F(Z^{\dag}) \, .
\label{cfF2}
\ee
The bosonic canonical commutation relations
that must be satisfied by the transformed
modes are
\be
[b_{i},b_{j}^{\dag}]=\delta_{ij} \;
\quad [b_{i},b_{j}]=0 \; .
\label{nlccr}
\ee
We now show that they are satisfied if some simple algebraic
constraints hold between the coefficients of the nonlinear
transformations. These algebraic relations generalize
Eq.~(\ref{lcc}) of the linear case, and are
independent of the particular form of the function $F$.
The first condition is again Eq.~(\ref{lcc}). It derives
from the $c$-number part of the canonical commutation relations
Eqs.~(\ref{nlccr}) which, in turn, is due to the linear part of
the transformations Eqs.~(\ref{nlct}).
The new conditions derive from the
operator-valued part of Eqs.~(\ref{nlccr}). Exploiting the
constraints on the nonlinear function $F$ given by
Eqs.~(\ref{cfF1})-(\ref{cfF2}), and using the
commutation rules for functions of the annihilation and
creation operators (see also Part I \cite{paper1}),
one finally has
\bea
&& |\alpha|^{2}-|\beta|^{2}=0 \, , \nonumber \\
\label{nlcc} \\
&& \mu\chi^{*}\alpha-\nu\chi^{*}\beta^{*}
+\mu^{*}\gamma\beta^{*}-\nu^{*}\gamma\alpha=0 \,. \nonumber
\eea
We see that conditions Eqs.~(\ref{nlcc}) are simple algebraic
relations on the coefficients of the transformations.
Conditions Eqs.~(\ref{lcc})-(\ref{nlcc}) are compatible with
very many specific canonical realizations of the nonlinear
transformations. To restrict further the possible canonical
choices, we will adopt the parameterizations Eq.~(\ref{lp}) and
\bea
{\quad \gamma} & = &
|\gamma| e^{i \delta_{1} } \;,
\; {\quad  \chi} =  |\chi| e^{i\delta_{2}} \;, \nonumber \\
\label{nlp} \\
{\quad \alpha} & = & |\alpha| e^{i \theta_{1} } \; \; \; ,
\;  {\quad \beta} = |\beta| e^{i \theta_{2}} \; . \nonumber
\eea
Let us finally choose
$|\alpha|=|\beta|=\frac{1}{\sqrt{2}}$, so that the
moduli (strengths) $|\gamma|$ and $|\chi|$ of the nonlinear
couplings, and the phases $\delta_{1}$, $\delta_{2}$, $\theta_{1}$,
and $\theta_{2}$ are the remaining free parameters.
The argument of the nonlinear function
$F$ in Eq.~(\ref{cfF1}) becomes
$(e^{-i\theta_{2}}a_{2}+
e^{i\theta_{1}}a^{\dag}_{1})/\sqrt{2}$, and it can be
interpreted as the output photocurrent of an ideal
heterodyne detector \cite{shapiro1,shapiro2}.
We see that the structure of the two-mode, multiphoton
canonical transformations is intrinsically based on
heterodyne mixings of the different field quadratures.
The transformations Eqs.~(\ref{nlct}) now read
\bea
&& b_{1}=\mu
a_{1} + \nu a_{2}^{\dag} + |\gamma|e^{i \delta_{1}}
F\left(\frac{e^{-i\theta_{2}}a_{2}+
e^{i\theta_{1}}a^{\dag}_{1}}{\sqrt{2}}\right) \; , \nonumber \\
&& \label{bop} \, \\
&& b_{2}=\mu a_{2} + \nu a_{1}^{\dag} + |\chi|e^{i \delta_{2}}
F\left(\frac{e^{-i\theta_{1}}a_{1}+
e^{i\theta_{2}}a^{\dag}_{2}}{\sqrt{2}}\right) \; , \nonumber
\eea
while Eqs.~(\ref{nlcc}) are encoded in the single condition
\bea
&& \cosh r |\chi| e^{-i(\delta_{2}-\theta_{1})}-\sinh r
|\chi|
e^{-i(\delta_{2}+\theta_{2}-\phi)} \nonumber \\
&&+\cosh r |\gamma|e^{i(\delta_{1}-\theta_{2})}-\sinh r |\gamma|
e^{i(\delta_{1}+\theta_{1}-\phi)}=0 \, . \nonumber \\
\label{nlcc1}
\eea
Eq.~(\ref{nlcc1}) can be written in the compact form
\begin{widetext}
\be
\tanh r=\frac{(|\chi|^{2}+|\gamma|^{2})\cos(\theta_{1}+\theta_{2}-\phi)
 +
2|\chi||\gamma|\cos(\delta_{1}+\delta_{2}-\phi)+ i
(|\chi|^{2}-|\gamma|^{2})\sin(\theta_{1}+\theta_{2}-\phi)}
{|\chi|^{2}+|\gamma|^{2}+2|\chi||\gamma|\cos(\delta_{1}+\delta_{2}+\theta_{1}+\theta_{2}-2\phi)
} \; .
\label{nlcc2}
\ee
\end{widetext}
Being $r$ real, the imaginary part of Eq.~(\ref{nlcc2})
must be set to zero:
\be
(|\chi|^{2}-|\gamma|^{2})\sin(\theta_{1}+\theta_{2}-\phi)=0
\, .
\label{imaginary}
\ee
Eq.~(\ref{imaginary}) is satisfied either if
\be
\theta_{1}+\theta_{2}-\phi=k \pi \, ,
\label{subc1}
\ee
or
\be
|\chi|=|\gamma| \, .
\label{subc2}
\ee
By choosing Eq.~(\ref{subc1}) then Eq.~(\ref{nlcc2}) reduces to
\be
\tanh r=\pm 1 \, , \nonumber
\ee
which corresponds to $r=\infty$.
By choosing instead Eq.~(\ref{subc2})
we have
\be
\tanh r=\frac{\cos(\theta_{1}+\theta_{2}-\phi)+
\cos(\delta_{1}+\delta_{2}-\phi)}
{1+\cos(\delta_{1}+\delta_{2}+\theta_{1}+\theta_{2}-2\phi) } \,
\label{nlcc22}
\ee
which is independent of the nonlinear strength $|\gamma|$
and can always be satisfied. The class of possible
solutions of Eq.~(\ref{nlcc22}) is rather large. In order
to find explicit examples we must specify some concrete realizations.
We then adopt the limiting solutions:
\be
\delta_{1}+\delta_{2}-\phi=0,\pm\pi \; \; , \; \;
\theta_{1}+\theta_{2}-\phi = \pm \pi ,0 \, .
\label{nlcc3}
\ee
The
transformations Eqs.~(\ref{bop}) define a diagonalized Hamiltonian
of the form Eq.~(\ref{gmH}). Inserting in this expression Eqs.~(\ref{bop})
for the transformed variables yields
multiphoton Hamiltonians written in terms of the fundamental variables
$a_{i}$ and $a_{i}^{\dag}$. They are parametrized
by $F$, and each one is characterized by a specific
choice of the nonlinear function.
Among all the many possible
forms we select some which bear particularly interesting and
realistic physical interpretations. Let us in fact consider
Hamiltonians associated to the following set of nonlinear functions
defined as $F(\zeta) = \zeta^n$, i.e. as integer powers of the
heterodyne variable. They describe nondegenerate and degenerate
processes up to $2n$-photon ones. In particular, we
concentrate our attention on the simplest choice of lowest
nonlinearity $F(\zeta) =
\zeta^2$, describing up to four-photon processes. The
Hamiltonian reads
\bea
&& H=A_{0}+B_{0}(a_{1}^{\dag}a_{1}+a_{2}^{\dag}a_{2}) \nonumber \\
&&+ C_{0}(a_{1}^{\dag 2}a_{1}^{2}+a_{2}^{\dag2}
a_{2}^{2}+2a_{1}^{\dag}a_{1}a_{2}^{\dag}a_{2}) \, \nonumber \\
&& +[D_{1}a_{1}^{\dag}a_{2}^{\dag}+D_{2}a_{1}^{\dag
2}a_{2}+D_{2}'a_{1}a_{2}^{\dag 2} \nonumber \\
&&+D_{3}a_{1}^{\dag
3}+D_{3}'a_{2}^{\dag 3}
+D_{4}a_{1}^{\dag 2}a_{2}^{\dag 2} \nonumber \\
&&+D_{5}(a_{1}^{\dag 2}a_{1}a_{2}^{\dag}+a_{1}^{\dag}a_{2}^{\dag
2}a_{2}) + h.c.] \, ,
\label{fpH}
\eea
where the coefficients are given by:
\bea
\quad A_{0} &=& |\gamma|^{2}+2|\nu|^{2} \, , \nonumber \\
\quad B_{0} &=&|\mu|^{2}+|\nu|^{2}+2|\gamma|^{2} \, , \nonumber \\
\quad C_{0} &=& \frac{1}{2}|\gamma|^{2} \, , \nonumber \\
\quad D_{1} &=& 2\mu^{*}\nu+2|\gamma|^{2}e^{i(\theta_{1}+\theta_{2})} \, ,
\nonumber \\
\quad D_{2} &=& |\gamma|e^{i\theta_{1}}
[\frac{1}{2}e^{i(\theta_{1}-\delta_{2})}\mu
+ e^{-i(\theta_{2}+\delta_{2})}\nu \, , \nonumber \\
&+&e^{-i(\theta_{2}-\delta_{1})}\mu^{*}+
\frac{1}{2}e^{i(\theta_{1}+\delta_{1})}\nu^{*}] \, , \nonumber \\
\quad D_{2}' &=& |\gamma|e^{i\theta_{2}}
[\frac{1}{2}e^{i(\theta_{2}-\delta_{1})}\mu+
e^{-i(\theta_{1}+\delta_{1})}\nu \nonumber \\
&+&e^{-i(\theta_{1}-\delta_{2})}\mu^{*}+
\frac{1}{2}e^{i(\theta_{2}+\delta_{2})}\nu^{*}] \, , \nonumber \\
\quad D_{3} &=& \frac{1}{2}|\gamma|e^{2i\theta_{1}}
(e^{i\delta_{1}}\mu^{*}+e^{-i\delta_{2}}\nu) \, , \nonumber \\
\quad D_{3}' &=&\frac{1}{2}|\gamma|e^{2i\theta_{2}}
(e^{i\delta_{2}}\mu^{*}+e^{-i\delta_{1}}\nu) \, , \nonumber \\
\quad D_{4} &=&
\frac{|\gamma|^{2}}{2}e^{2i(\theta_{1}+\theta_{2})} \, , \; \;
D_{5}=|\gamma|^{2}e^{i(\theta_{1}+\theta_{2})} \, .
\label{coeff}
\eea
The parameters are constrained by the canonical
conditions Eqs.~(\ref{nlcc3}). Regarding $A_{0}, B_{0}, C_{0}$, it
is sufficient to insert the parametrization Eqs.~(\ref{lp}). To be
concrete, for the remaining parameters we take the
canonical choice $\delta_{1} + \delta_{2} - \phi = 0$ and
$\theta_{1}+\theta_{2}-\phi=\pi$, obtaining
\bea
&&D_{1}=e^{i \phi}(\sinh 2r-2|\gamma|^{2}) \, , \nonumber \\
&& \nonumber \\
&&D_{2}=-\frac{|\gamma|}{2}e^{r}e^{i (2\theta_{1}-\delta_{2})}
\, , \,
\quad D_{2}'=-\frac{|\gamma|}{2}e^{r}e^{i (2\theta_{2}-\delta_{1})}
\, , \nonumber \\
&& \nonumber \\
&&D_{3}=\frac{|\gamma|}{2}e^{r}e^{i (2\theta_{1}+\delta_{1})}
\, , \,
\quad D_{3}'=\frac{|\gamma|}{2}e^{r}e^{i (2\theta_{2}+\delta_{2})}
\, , \, \nonumber \\
&& \nonumber \\
&&D_{4}=\frac{|\gamma|^{2}}{2}e^{2i\phi} \, ,\, \quad
D_{5}=-|\gamma|^{2}e^{i \phi} \, .
\label{coeff1}
\eea
As previously stated, the Hamiltonian Eq.~(\ref{fpH}) describes up to
four-photon degenerate and non degenerate processes.
We conclude this Section noting that for degenerate, single-mode
processes, the heterodyne canonical formalism reduces to the
homodyne canonical formalism introduced in Part I \cite{paper1}.

\section{Entangled state representation}
\label{entstrep}

In this and in the following Sections we determine
the form and the statistical properties of the coherent states
associated to the heterodyne canonical transformations, at least in
the simplest cases.
In order to do this, it is convenient to introduce preliminarily
the so called {\it entangled state representation}
\cite{hong1,hong2,dariano,hong3}.

We define the non Hermitian operators :
\be
Z=\frac{e^{-i\theta_{2}}a_{2}+
e^{i\theta_{1}}a^{\dag}_{1}}{\sqrt{2}} \,  \quad
P_{z}=i\frac{e^{i\theta_{2}}a^{\dag}_{2}-
e^{-i\theta_{1}}a_{1}}{\sqrt{2}} \, ,
\label{ZPz}
\ee
which satisfy the commutation relations
\be
[Z,Z^{\dag}]=[P_{z},P_{z}^{\dag}]=0, \quad [Z,P_{z}]=i \, .
\label{crZPz}
\ee
Note that $Z$ is just the argument of the
nonlinear function $F$ in the transformations Eqs.~(\ref{bop}): as
already mentioned, this operator can be interpreted as the output
photocurrent of an ideal heterodyne detector \cite{shapiro1,shapiro2}.

In terms of $Z$, $P_{z}$, Eqs.~(\ref{bop}) become
\bea
&& b_{1}=\frac{\mu'+\nu''}{\sqrt{2}}Z^{\dag}+i
\frac{\mu'-\nu''}{\sqrt{2}}P_{z}+|\gamma|e^{i\delta_{1}}F(Z)
\,, \nonumber \\ && \label{bopZPz} \,  \\
&& b_{2}=\frac{\mu''+\nu'}{\sqrt{2}}Z+i
\frac{\mu''-\nu'}{\sqrt{2}}P_{z}^{\dag}+
|\gamma|e^{i\delta_{2}}F(Z^{\dag})
\, , \nonumber
\eea
where $\mu'=e^{i\theta_{1}}\mu$,
$\mu''=e^{i\theta_{2}}\mu$,  $\nu'=e^{-i\theta_{1}}\nu$,
$\nu''=e^{-i\theta_{2}}\nu$.
The algebra defined by relations Eqs.~(\ref{crZPz}) and
(\ref{ZPz}) is characterized by the orthonormal eigenvectors
$\{|z\rangle\}$ of $X_{\theta_{1}}+X_{\theta_{2}}$ and
$P_{\theta_{2}}-P_{\theta_{1}}$ \cite{hong1}, also called
entangled-state representation, where $X_{\theta_{i}}$ and
$P_{\theta_{i}}\equiv X_{\theta_{i}+\frac{\pi}{2}}$, $i=1,2$, are
the homodyne quadrature operators for the mode $a_i$, and $z$ is an
arbitrary complex number $z=z_{1} + iz_{2}$. It can be easily
proved that in the two-mode Fock space the state $|z\rangle$ is
\bea
|z\rangle &=& \exp\left[-|z|^{2}+\sqrt{2}z
a_{\theta_{2}}^{\dag}+\sqrt{2}z^{*} a_{\theta_{1}}^{\dag}-
a_{\theta_{1}}^{\dag}a_{\theta_{2}}^{\dag}\right]|00\rangle
\nonumber \\
&& \nonumber \\
& = & \exp\left[\sqrt{2}z a_{\theta_{2}}^{\dag}-\sqrt{2}z^{*}
a_{\theta_{2}}\right]
\exp\left[-a_{\theta_{1}}^{\dag}a_{\theta_{2}}^{\dag}\right]|00
\rangle ,
\label{zstate}
\eea
where $|00\rangle$ is the two--mode vacuum. The
states Eq.~(\ref{zstate}) satisfy the eigenvalue equations
\be
(X_{\theta_{1}}+X_{\theta_{2}})|z\rangle =2z_{1}|z\rangle \; \; ,
(P_{\theta_{2}}-P_{\theta_{1}})|z\rangle =2z_{2}|z\rangle \, ,
\ee
and
\be
Z|z\rangle=z|z\rangle \, \quad \quad
Z^{\dag}|z\rangle=z^{*}|z\rangle \, .
\label{zeieq}
\ee
The states $\{|z\rangle\}$ are orthonormal
\be
\langle z'|z\rangle=\pi \delta^{(2)}(z'-z) \, ,
\ee
and satisfy the
completeness relation
\bea
&&\frac{2}{\pi}\int d^{2}z
|z\rangle\langle z| = \nonumber \\
&&\frac{2}{\pi}\int d^{2}z
:\exp\{-2|z|^{2}+\sqrt{2}z(a_{\theta_{2}}^{\dag}+a_{\theta_{1}})
\nonumber \\
&&+\sqrt{2}z^{*}(a_{\theta_{1}}^{\dag}+a_{\theta_{2}})-
(a_{\theta_{2}}^{\dag}+
a_{\theta_{1}})(a_{\theta_{1}}^{\dag}+a_{\theta_{2}})\}:
\; = \; 1 \, , \nonumber \\
\label{zcompl}
\eea
where $d^{2}z \equiv dz dz^{*}$.
In the next Section we will exploit the entangled
state representation to determine the coherent states
associated to the nonlinear canonical transformations.

\section{HEMPSS: Heterodyne multiphoton squeezed states}
\label{eigenveq}

As is well known, the Glauber (harmonic oscillator) coherent
states can be defined via three equivalent procedures
\cite{Klauder}. For Hamiltonians whose elements belong to more complex
algebras than the Weyl--Heisenberg algebra, the three procedures
are in general not equivalent, and lead to different definitions of
coherent states \cite{Klauder,gencoh}. The coherent states
associated to the unitary Hamiltonian evolution applied to the ground
state are generated by acting with the displacement operator on the
vacuum. This is not our case: as anticipated in the
Introduction, we have been faced with the question of constructing operator
variables which form a harmonic-oscillator
Weyl--Heisenberg algebra, although containing nonlinear terms.
Having solved the problem in the entangled state representation,
we can now determine the coherent states defined as simultaneous
eigenvectors of the transformed operators Eqs.~(\ref{bop}).
The associated eigenvalue equations are
\begin{equation}
b_{i}|\psi\rangle_{\beta}=\beta_{i}|\psi\rangle_{\beta}  \; ,
\; \; \; i=1,2 \, ,
\end{equation}
which in the $\{|z\rangle\}$ representation become
\begin{equation}
\langle z|b_{i}|\psi\rangle_{\beta}=\beta_{i}\langle
z|\psi\rangle_{\beta} \; .
\label{zed}
\end{equation}
Exploiting Eqs.~(\ref{bopZPz}) and the differential
representation for the operators $\{Z,P_{z}\}$, we can write
Eqs.~(\ref{zed}) in the differential form
\bea
&& \frac{\mu'+\nu''}{\sqrt{2}}z\psi_{\beta}
+\frac{\mu'-\nu''}{\sqrt{2}}\partial_{z^{*}}\psi_{\beta}+
|\gamma|e^{i\delta_{1}}F(z^{*})\psi_{\beta}=
\beta_{1}\psi_{\beta} \, , \nonumber \\
&& \\
&& \frac{\mu''+\nu'}{\sqrt{2}}z^{*}\psi_{\beta}
+\frac{\mu''-\nu'}{\sqrt{2}}\partial_{z}\psi_{\beta}+
|\gamma|e^{i\delta_{2}}F(z)\psi_{\beta}= \beta_{2}\psi_{\beta} \, ,
\nonumber
\eea
whose solution is
\be
\psi_{\beta}(z,z^{*})=N
e^{-a|z|^{2}+\Gamma_{1}z^{*}+\Gamma_{2}z- B(z,z^{*})} \, .
\label{wf}
\ee
In Eq.~(\ref{wf}), $N$ is a normalization factor, the function
$B(z,z^{*})$ (which encodes the nonlinear contribution) is
\be
B(z,z^{*})=b_{1}\int^{z^{*}}d\xi^{*}F(\xi^{*})+b_{2}\int^{z}d\xi
F(\xi) \, ,
\ee
and
\bea
&&a=\frac{\mu'+\nu''}{\mu'-\nu''}=\frac{\mu''+\nu'}{\mu''-\nu'}
=\frac{1+2i Im(\mu'^{*}\nu'')}{|\mu'-\nu''|^{2}} \, , \nonumber
\\ && \nonumber \\
&&b_{1}=\frac{\sqrt{2}|\gamma|e^{i\delta_{1}}}{\mu'-\nu''} \, , \,
\quad
b_{2}=\frac{\sqrt{2}|\gamma|e^{i\delta_{2}}}{\mu''-\nu'} \, , \\
&& \nonumber \\
&&\Gamma_{1}=\frac{\sqrt{2}\beta_{1}}{\mu'-\nu''} \, , \, \quad
\Gamma_{2}=\frac{\sqrt{2}\beta_{2}}{\mu''-\nu'} \, . \nonumber
\eea
We remark that the canonical conditions Eqs.~(\ref{nlcc3}) imply
$Re[a]>0$, $Im[a]=0$ and $Re[B(z,z^{*})]=0$, which
ensure the normalizability of the wave function Eq.~(\ref{wf}).
The states Eq.~(\ref{wf}) satisfy the (over)completeness
relation
\be
\frac{1}{\pi^{2}}\int
d^{2}\beta_{1}d^{2}\beta_{2} |\psi\rangle_{\mathbf{\beta}}
\langle\psi| = 1 \, .
\ee
Of course, for $\gamma=0$ the states
Eq.~(\ref{wf}) reduce to the standard two-mode squeezed
states. The states Eq.~(\ref{wf}) are two-mode multiphoton
states, depending on heterodyne combinations of the field
modes. They share properties of coherence and squeezing,
as we will show in more detail in the following. We thus
name them heterodyne multiphoton squeezed states (HEMPSS).
For a single mode, they reduce to the homodyne multiphoton
squeezed states (HOMPSS) introduced in the companion
paper Part I.
We have explicitly obtained the general form of the wave
functions of the HEMPSS in the entangled state
representation. The general properties of these states, which we
will study in the following Sections, are however easier to
investigate in the coordinate representation. Let us first
recall the following useful expression for $|z\rangle$
\cite{hong2,dariano}:
\be
|z\rangle=e^{-2i z_{1}z_{2}}
\int_{-\infty}^{\infty}d\xi e^{2i \xi z_{2}} |2
z_{1}-\xi\rangle_{\theta_{1}} \otimes|\xi\rangle_{\theta_{2}} \, ,
\label{coordrep}
\ee
where the tensor product $|\psi\rangle
\otimes |\phi\rangle$ denotes kets in the Hilbert space
$H_{1}\otimes H_{2}$, and $|\xi\rangle_{\theta_{i}}$ is an eigenvector
of the quadrature operator $X_{\theta_{i}}$ of the
$i$--th mode ($i=1,2$).
From Eq.~(\ref{coordrep}) it follows
\be
\psi_{\beta}(x_{\theta_{1}},x_{\theta_{2}}) =
\frac{2}{\pi}\int_{-\infty}^{\infty}d z_{2}e^{2i (x_{\theta_{2}}
-x_{\theta_{1}})z_{2}} \psi_{\beta} (z_1, z_2) \, ,
\label{wfxy}
\ee
where $\psi_{\beta}$ is evaluated at $z_1 =
\frac{x_{\theta_{1}} + x_{\theta_{2}}}{2}$.
The states Eq.~(\ref{wfxy}) are in rather implicit form. Under
particular conditions one can however find explicit analytic
expressions. For instance, considering a quadratic
nonlinearity $F(Z)=Z^{2}$, and letting
$\delta_{1}-\theta_{1}=\frac{\pi}{2}$, they take the form
\bea
&&
\psi_{\beta}(x_{\theta_{1}},x_{\theta_{2}})=
\frac{2}{\sqrt{\pi}}\frac{N}{\sqrt{a-\frac{3}{2}i\Xi(x_{\theta_{1}}
+x_{\theta_{2}})}} \nonumber \\
&&\cdot \exp\{-\frac{[(x_{\theta_{1}}-x_{\theta_{2}})
+(\Gamma_{1}-\Gamma_{2})]^{2}}{4a-6i\Xi(x_{\theta_{1}}
+x_{\theta_{2}})}\} \, \nonumber \\
&&\cdot \exp\{-i\Xi(\frac{x_{\theta_{1}}+x_{\theta_{2}}}{2})^{3}
-a(\frac{x_{\theta_{1}}+x_{\theta_{2}}}{2})^{2} \nonumber \\
&&+(\Gamma_{1}+\Gamma_{2})(\frac{x_{\theta_{1}}+x_{\theta_{2}}}{2})\}
\, ,
\label{explstate}
\eea
where, due to the canonical conditions,
$$
\Xi=\frac{2\sqrt{2}}{3}|\gamma|\frac{\cosh r-\sinh r
e^{i(\theta_{1}+\theta_{2}-\phi)}}{\cosh 2r-\sinh 2r \cos
(\theta_{1}+\theta_{2}-\phi)}
$$
is a real number.

\section{Unitary Operators}
\label{Uop}

The heterodyne multiphoton squeezed states (HEMPSS)
defined in the preceding Section
are not the states unitarily evolved by the
Hamiltonians associated to the nonlinear canonical transformations.
As in the degenerate case, however, they can be
unitarily generated from the two--mode
vacuum. By looking at the entangled state representation
of the HEMPSS, Eq.~(\ref{wf}), it is evident that
\be
|\psi\rangle_{\beta}=U(Z,Z^{\dag})D_{1}(\alpha_{1})
D_{2}(\alpha_{2})S_{12}(g)|00\rangle \, ,
\label{unitoptot}
\ee
where
\bea
&&D_{i}(\alpha_{i})=e^{\alpha_{i}a_{i}^{\dag}-\alpha_{i}^{*}a_{i}}
\, \nonumber \\
&&S_{12}(g)=e^{-g
a_{\theta_{1}}^{\dag}a_{\theta_{2}}^{\dag}+g^{*}a_{\theta_{1}}a_{\theta_{2}}}
\, ,
\eea
with $g=\pm r$ and $\alpha_{i}=\mu^{*}\beta_{i}-\nu\beta_{j}^{*}$
($i\neq j =1,2$). These two operators generate a two-mode
squeezed state from the two-mode vacuum.
The operator $U$ takes the form
\be
U(Z,Z^{\dag})=\exp\left\{-b_{1}\int^{Z}d\xi
F(\xi)-b_{2}\int^{Z^{\dag}}d\xi F(\xi)\right\} \, ,
\label{Uphas}
\ee
and the canonical conditions assure that it is unitary.
Obviously, for $\gamma=0$, the state $|\psi\rangle_{\beta}$ reduces
to an ordinary two--mode squeezed state.
In the case of lowest nonlinearity $F(Z)=Z^{2}$, exploiting
the canonical conditions, the operator Eq.~(\ref{Uphas}) reads
\be
U(Z,Z^{\dag})=e^{-\Delta Z^{3}+\Delta^{*}Z^{\dag 3}} \,
\label{Uph3} \, ,
\ee
where
\be
\Delta=\frac{\sqrt{2}|\gamma|e^{i(\delta_{1}-\theta_{1})}}{3(\mu-\nu
e^{-i(\theta_{1}+\theta_{2})})} =
\frac{\sqrt{2}e^{-r}|\gamma|}{3}e^{i(\delta_{1}-\theta_{1})} \, .
\label{nonlinstrength}
\ee
In this expression the exponent of the
operator Eq.~(\ref{Uph3}) is clearly anti-hermitian.
Depending on the specific canonical choice, the
parameter $r$ can acquire both positive and negative values.
It is interesting to express the operator Eq.~(\ref{Uph3}) in terms
of the rotated modes $a_{\theta_{i}}$ and $a_{\theta_{i}}^{\dag}$:
\be
U=\exp\left\{-\frac{\Delta}{2\sqrt{2}}[a_{\theta_{1}}^{\dag
3}+3a_{\theta_{1}}^{\dag
2}a_{\theta_{2}}+3a_{\theta_{1}}^{\dag}a_{\theta_{2}}^{2}+
a_{\theta_{2}}^{3}]+h.c.\right\} \, .
\label{Uf2}
\ee
In Section~\ref{expsetup} we will use this form to propose
a possible experimental realization of the HEMPSS.
As it is evident by their expressions, the HEMPSS
are non Gaussian, entangled states of bipartite
systems.

\section{Photon statistics}
\label{stat}

In this Section we study the statistics of the
four--photon HEMPSS. We first compute the two--mode
photon number distribution (PND). We compare it
with the PND of the two--mode squeezed states, thoroughly
studied in Ref.~\cite{PND}, and we show its strong
dependence on the modulus and phases of the nonlinear
couplings, and on the heterodyne mixing angles. We next show
the behavior of the average photon numbers in the two modes
as functions of $|\gamma |$. We then look at the second order
correlation functions, whose behaviors are strongly
dependent on the parameters of the nonlinear transformations
as well.

\begin{figure}
\begin{center}
\includegraphics*[width=8cm]{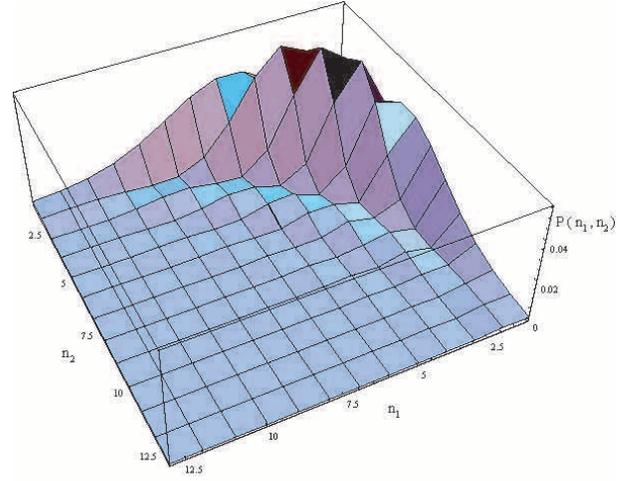}
\end{center}
\caption{$P(n_{1},n_{2})$ for a two--mode squeezed state, for
$r=0.8$, $\beta_{1}=\beta_{2}=3$ and $|\gamma|=0$.}
\label{pnss1}
\end{figure}

\begin{figure}
\begin{center}
\includegraphics*[width=8cm]{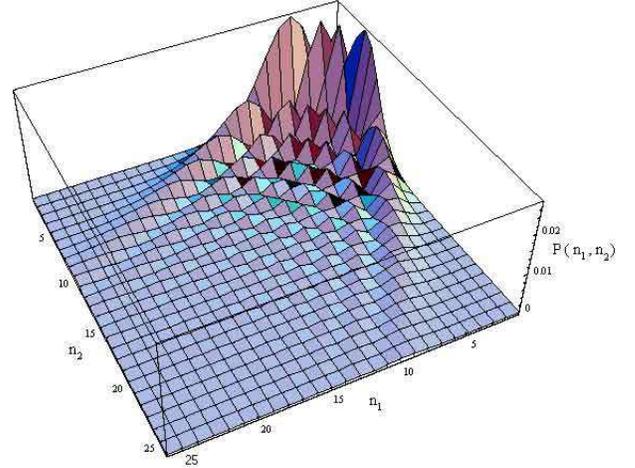}
\end{center}
\caption{$P(n_{1},n_{2})$ for two--mode squeezed state, for
$r=1.5$, $\beta_{1}=\beta_{2}=5$ and $|\gamma|=0$.}
\label{pnss2}
\end{figure}

Let us consider the scalar product $\langle
n_{1},n_{2}|z\rangle$
\bea
&&\langle n_{1},n_{2}|z\rangle=
(-1)^{m} e^{i(n_{1}\theta_{1}+n_{2}\theta_{2})}2^{(M-m)/2}e^{-|z|^{2}}
\left(\frac{m!}{M!}\right)^{1/2} \nonumber \\
&&\cdot z^{* n_{1}-m}z^{n_{2}-m}
L_{m}^{(M-m)}(2|z|^{2}) \, .
\label{zn}
\eea
Here $m=min(n_{1},n_{2})$, $M=max(n_{1},n_{2})$ and
$L_{n}^{\alpha}(x)$ are the generalized Laguerre polynomials.
Eq.~(\ref{zn}), together with the relation of
completeness Eq.~(\ref{zcompl}),
allows to write the following integral expression for the photon
number distribution (PND):
\bea
&& P(n_{1},n_{2})=|\langle
n_{1},n_{2}|\psi\rangle_{\beta}|^{2}= \frac{2^{M-m+2}}{\pi^{2}}
\left(\frac{m!}{M!}\right) \nonumber \\
\label{PND} \\
&&\cdot \left|\int d^{2}z e^{-|z|^{2}} z^{* n_{1}-m}z^{n_{2}-m}
L_{m}^{(M-m)}(2|z|^{2})\psi_{\beta}(z)\right|^{2} \, . \nonumber
\eea
This expression can be computed numerically, and plotted for
different values of the parameters. We compare the
four--photon HEMPSS with the standard two--mode
squeezed states. The case $\gamma = 0$ yields the PND computed
in Ref.~\cite{PND}. In Fig.~\ref{pnss1} the plot is drawn for
an intermediate value of the
squeezing parameter ($r = 0.8$), and a low mean number of photons,
while in Fig.~\ref{pnss2} we take a strong value of the squeezing
parameter ($r = 1.5$) at a larger mean number of photons. The PND is
symmetric with respect to $n_{1}$ and $n_{2}$, and exhibits the
characteristic oscillations due to interference in phase space
\cite{schleich}; the oscillations are enhanced for higher
squeezing.
We have plotted the two-mode,
four--photon PND for different values of the parameters.
In all graphs we have kept the same values of $r$, and the
corresponding values of $\beta_1 \; , \; \beta_2$
used for the standard squeezed states
($r = 0.8 \; , \; \beta_1 = \beta_2 = 3 \; ; \;
r = 1,5 \; , \; \beta_1 = \beta_2 = 5$).
We first set to zero the heterodyne mixing angles and
the phase of the squeezing: $\theta_1 = \theta_2 = \phi = 0$.
This choice forces $\delta_{1} + \delta_{2} = \pi$ (see Eq.~(\ref{nlcc3})).
Among all the possible realizations of this constraint we have
selected the completely symmetric choice $\delta_{1} = \delta_{2} = \pi/2$,
and the completely asymmetric one
$\delta_{1} = \pi \; , \; \delta_{2} = 0$.
For each value of $|\gamma |$
we have then four plots for different values
of the squeezing and of the phases. We choose, respectively,
$|\gamma | = 0.1$ and $|\gamma | = 0.2$.

\begin{figure}
\begin{center}
\includegraphics*[width=8cm]{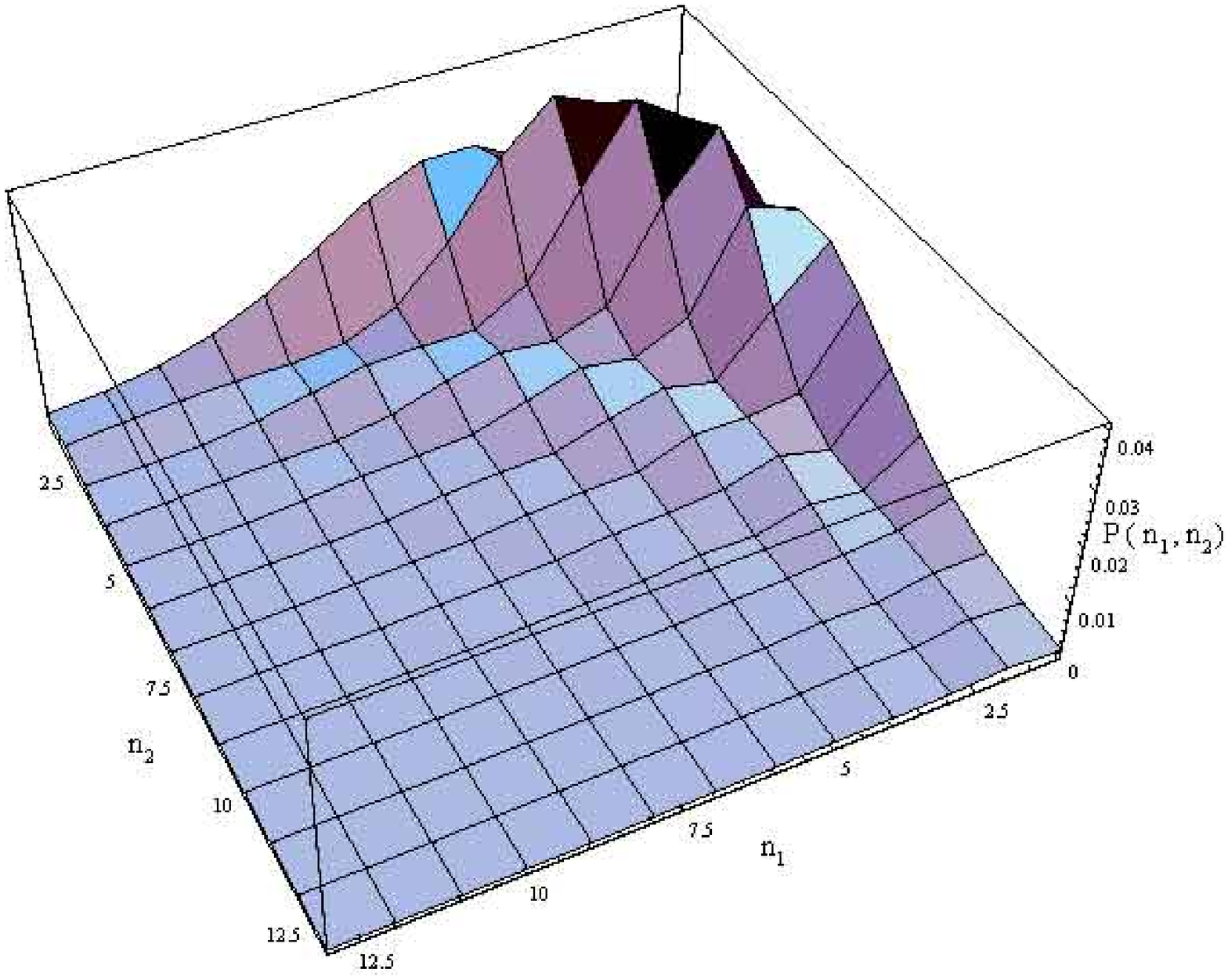}
\end{center}
\caption{$P(n_{1},n_{2})$ of the four-photon HEMPSS, for $r=0.8$,
$\beta_{1}=\beta_{2}=3$ and $|\gamma|=0.1$,
$\theta_{1}=\theta_{2}=0$, $\delta_{1}=\delta_{2}=\frac{\pi}{2}$.}
\label{pnhms1}
\end{figure}

\begin{figure}
\begin{center}
\includegraphics*[width=8cm]{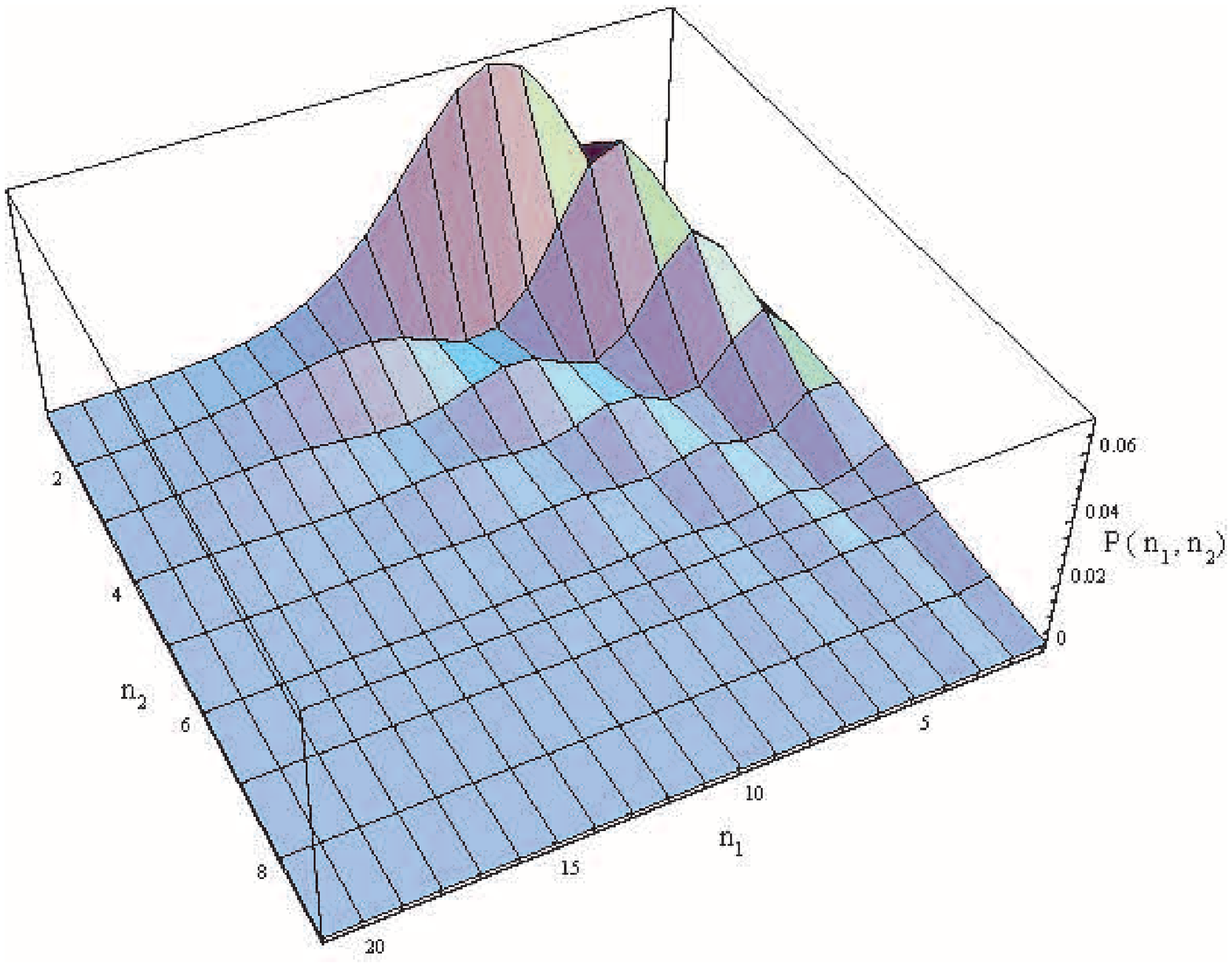}
\end{center}
\caption{$P(n_{1},n_{2})$ of the four-photon HEMPSS, for $r=0.8$,
$\beta_{1}=\beta_{2}=3$ and $|\gamma|=0.1$,
$\theta_{1}=\theta_{2}=0$, $\delta_{1}=\pi$, $\delta_{2}=0$.}
\label{pnhms2}
\end{figure}

\begin{figure}
\begin{center}
\includegraphics*[width=8cm]{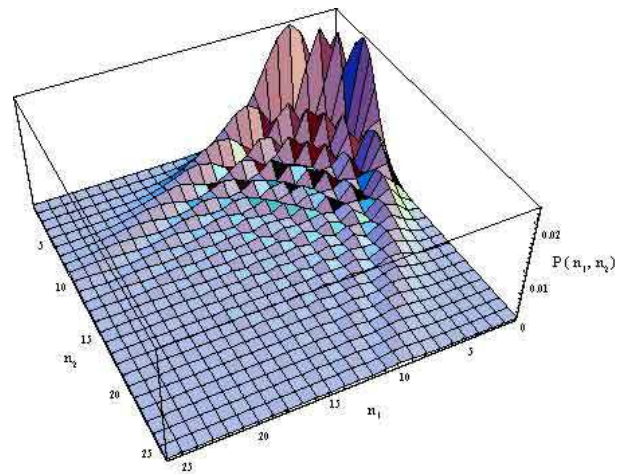}
\end{center}
\caption{$P(n_{1},n_{2})$ of the four-photon HEMPSS, for $r=1.5$,
$\beta_{1}=\beta_{2}=5$ and $|\gamma|=0.1$,
$\theta_{1}=\theta_{2}=0$, $\delta_{1}=\delta_{2}=\frac{\pi}{2}$.}
\label{pnhms3}
\end{figure}

\begin{figure}
\begin{center}
\includegraphics*[width=8cm]{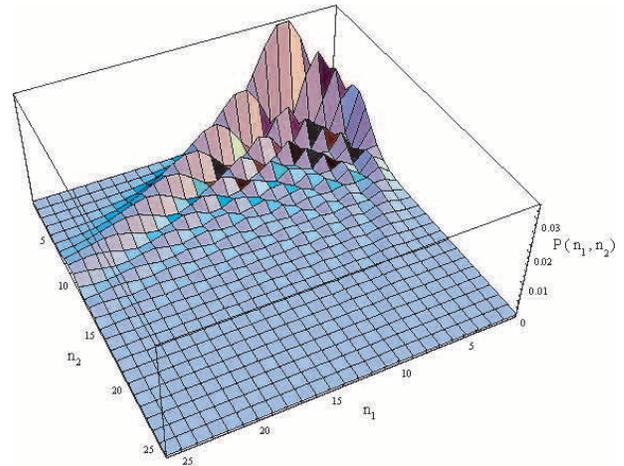}
\end{center}
\caption{$P(n_{1},n_{2})$ of the four-photon HEMPSS, for $r=1.5$,
$\beta_{1}=\beta_{2}=5$ and $|\gamma|=0.1$,
$\theta_{1}=\theta_{2}=0$, $\delta_{1}=\pi$, $\delta_{2}=0$.}
\label{pnhms4}
\end{figure}

\begin{figure}
\begin{center}
\includegraphics*[width=8cm]{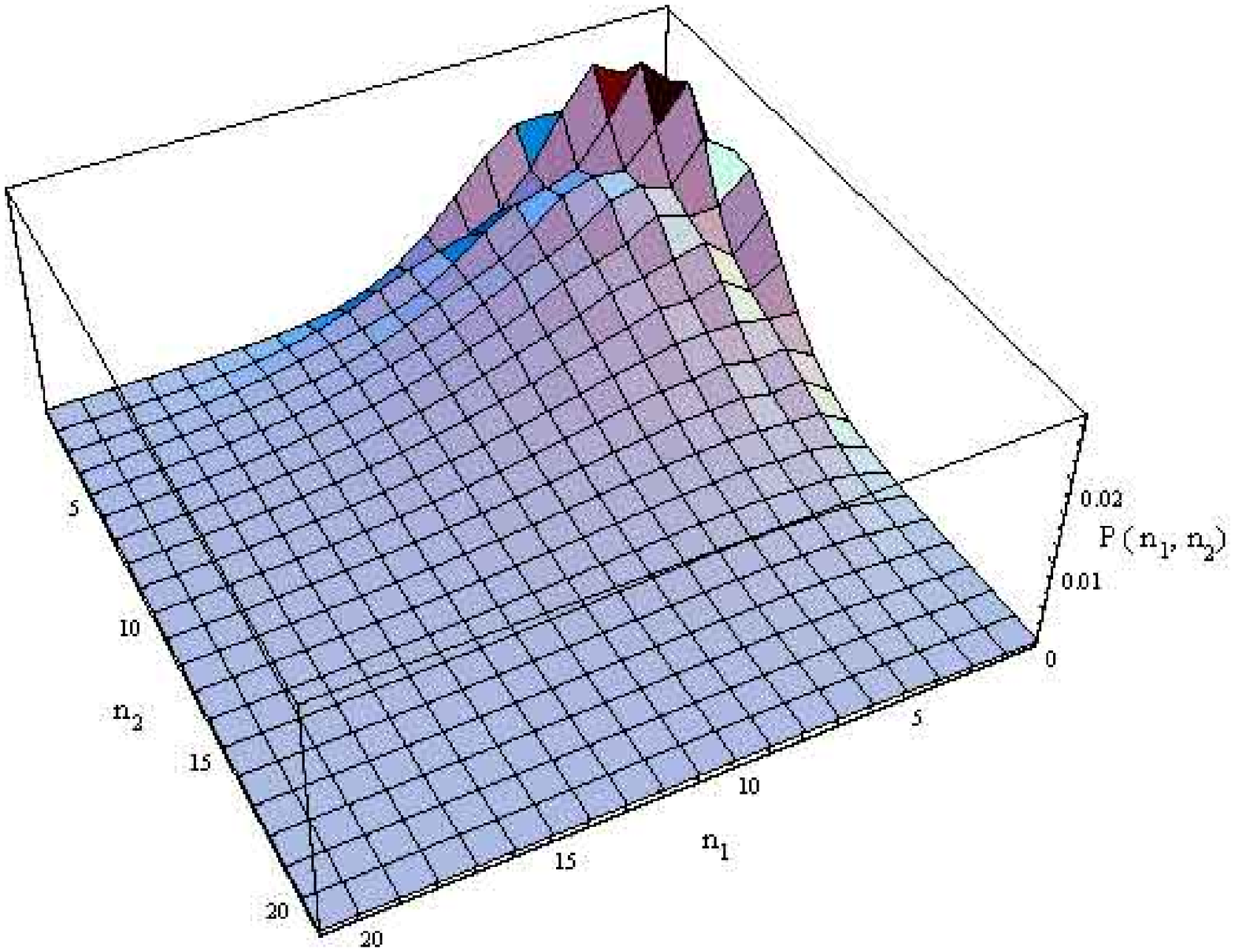}
\end{center}
\caption{$P(n_{1},n_{2})$ of the four-photon HEMPSS, for $r=0.8$,
$\beta_{1}=\beta_{2}=3$ and $|\gamma|=0.2$,
$\theta_{1}=\theta_{2}=0$, $\delta_{1}=\delta_{2}=\frac{\pi}{2}$.}
\label{pnhms5}
\end{figure}

\begin{figure}
\begin{center}
\includegraphics*[width=8cm]{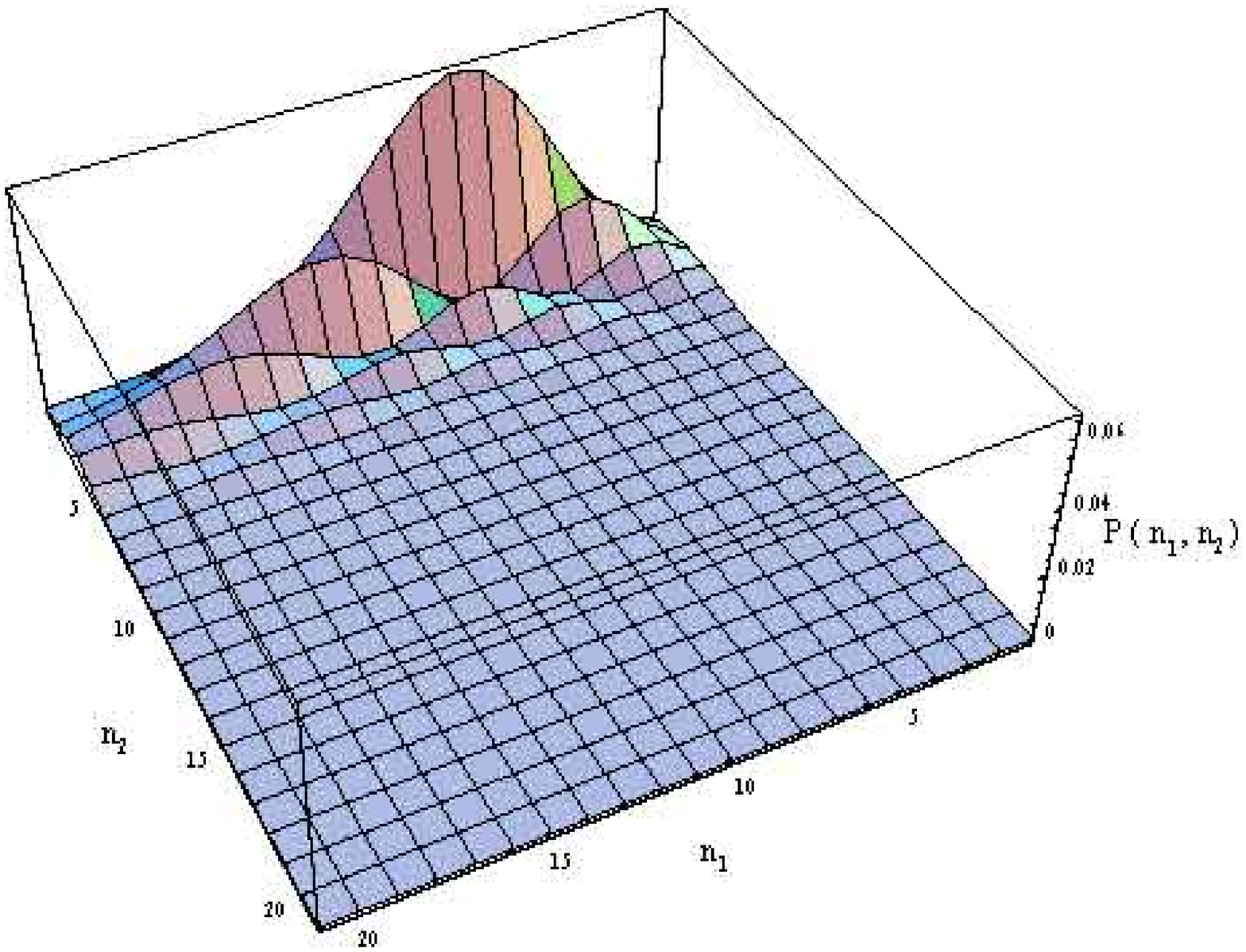}
\end{center}
\caption{$P(n_{1},n_{2})$ of the four-photon HEMPSS, for $r=0.8$,
$\beta_{1}=\beta_{2}=3$ and $|\gamma|=0.2$,
$\theta_{1}=\theta_{2}=0$, $\delta_{1}=\pi$, $\delta_{2}=0$.}
\label{pnhms6}
\end{figure}

\begin{figure}
\begin{center}
\includegraphics*[width=8cm]{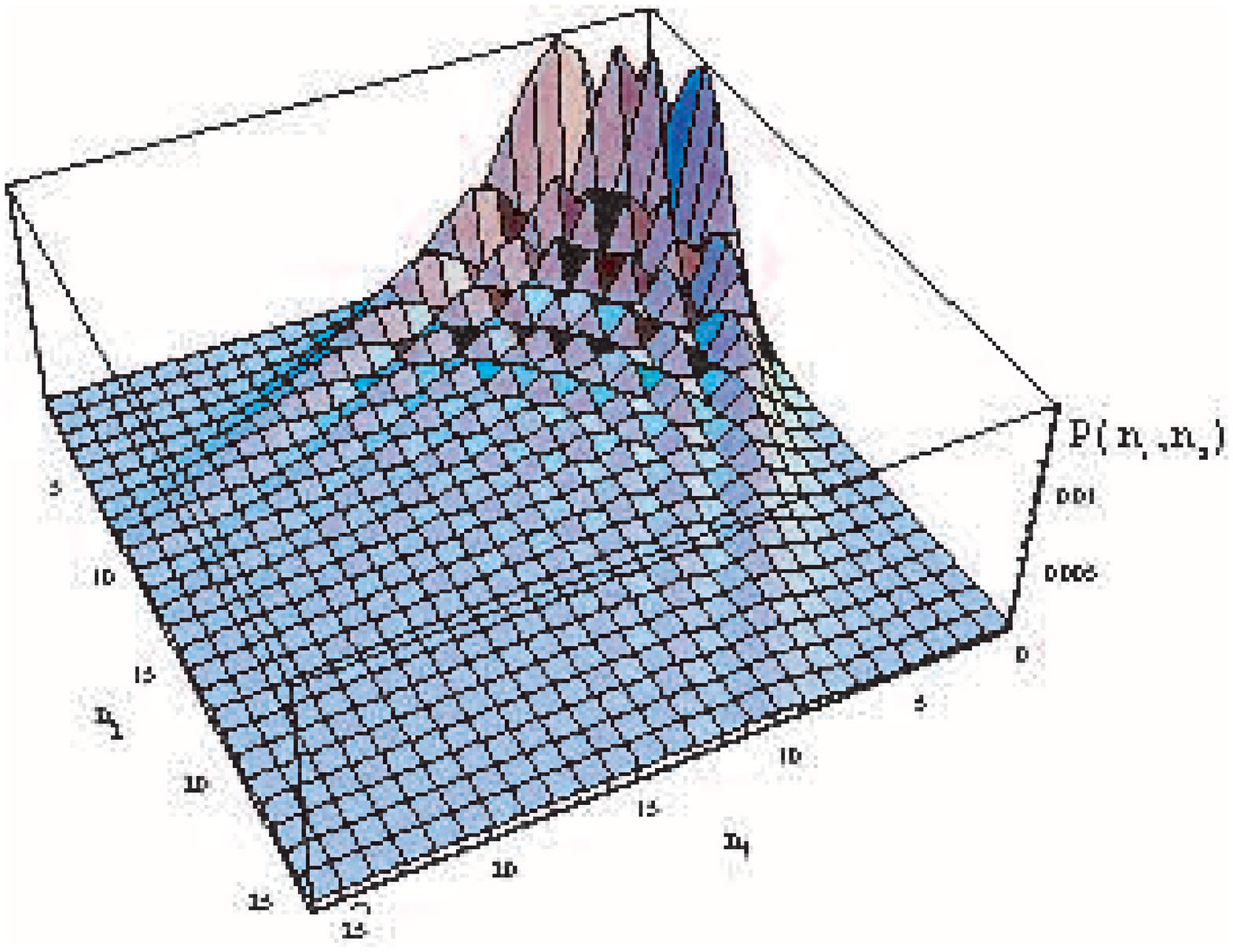}
\end{center}
\caption{$P(n_{1},n_{2})$ of the four-photon HEMPSS, for $r=1.5$,
$\beta_{1}=\beta_{2}=5$ and $|\gamma|=0.2$,
$\theta_{1}=\theta_{2}=0$, $\delta_{1}=\delta_{2}=\frac{\pi}{2}$.}
\label{pnhms7}
\end{figure}

\begin{figure}
\begin{center}
\includegraphics*[width=8cm]{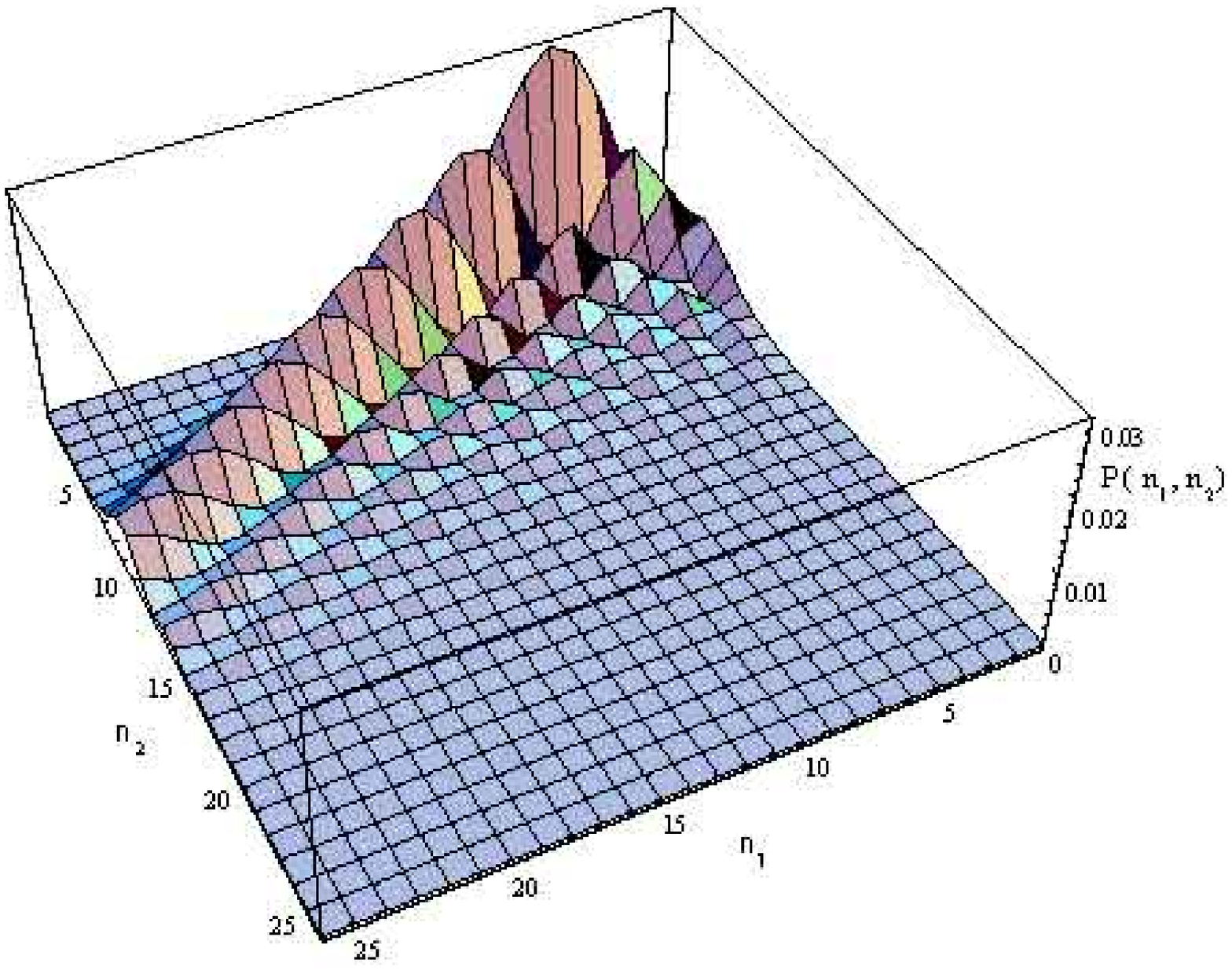}
\end{center}
\caption{$P(n_{1},n_{2})$ of the four-photon HEMPSS, for $r=1.5$,
$\beta_{1}=\beta_{2}=5$ and $|\gamma|=0.2$,
$\theta_{1}=\theta_{2}=0$, $\delta_{1}=\pi$, $\delta_{2}=0$.}
\label{pnhms8}
\end{figure}

We see that the form and the oscillations of the PND for
the four-photon HEMPSS strongly
depend on the values of the phases $\delta_{1}$ and
$\delta_{2}$ which produce competing effects in $U(Z, Z^{\dag})$. In fact, in
Figs.~\ref{pnhms1}, \ref{pnhms3}, \ref{pnhms5}, \ref{pnhms7},
corresponding to the symmetric choice for the phases, the PND is
symmetric with respect to $n_1$ and $n_2$. Viceversa, in
Figs.~\ref{pnhms2}, \ref{pnhms4}, \ref{pnhms6}, \ref{pnhms8} the
asymmetric choice of the phases leads to an asymmetric PND: the peaks are
displaced, and their number and form are changed.

\vspace{1cm}

\begin{figure}
\begin{center}
\includegraphics*[width=8cm]{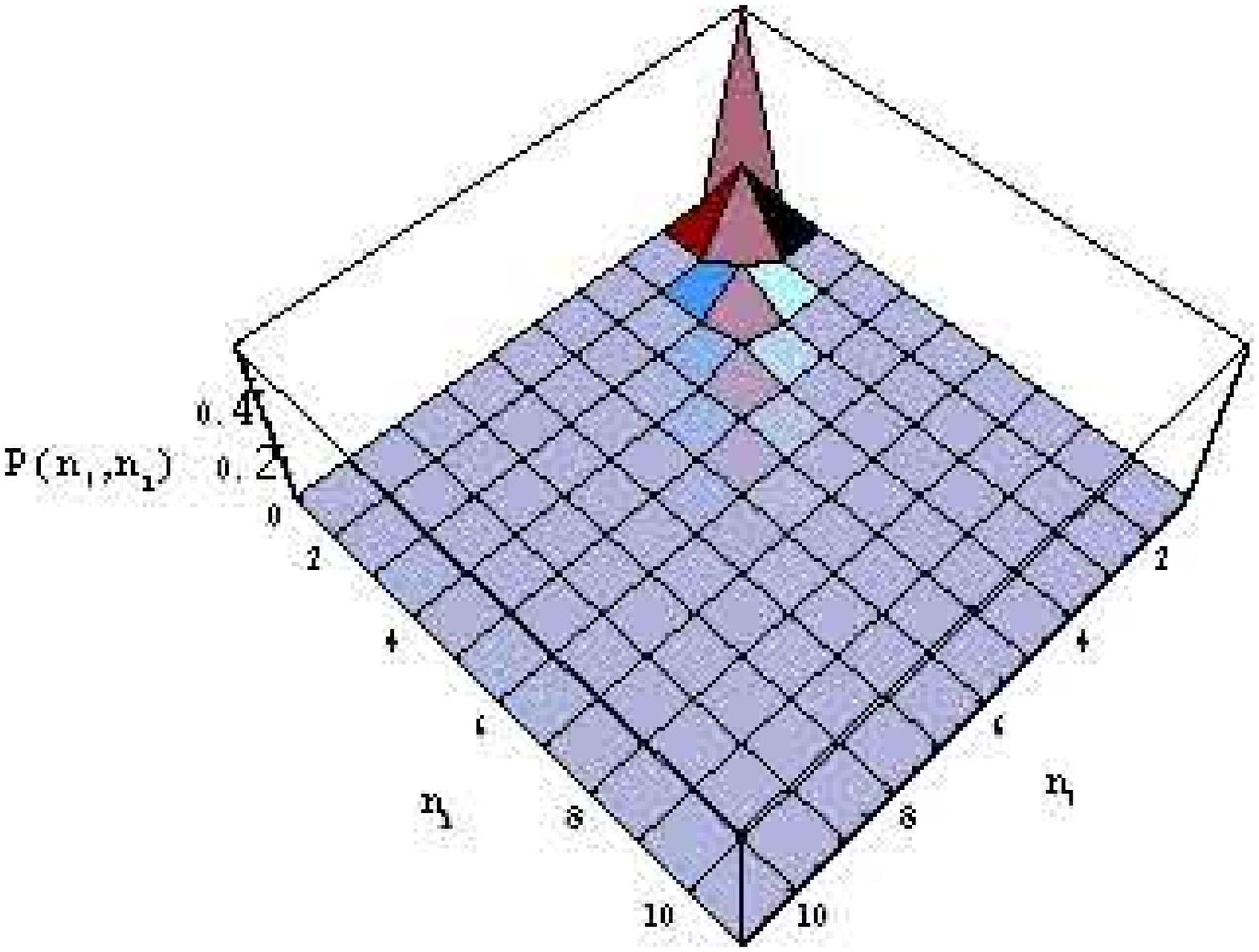}
\end{center}
\caption{$P(n_{1},n_{2})$ of the four-photon HEMPSS, for $r=0.8$,
$\beta_{1}=\beta_{2}=0$ and $|\gamma|=0.1$,
$\theta_{1}=\theta_{2}=0$, $\delta_{1}=\pi$, $\delta_{2}=0$.}
\label{pnhms10_2}
\end{figure}

\begin{figure}
\begin{center}
\includegraphics*[width=8cm]{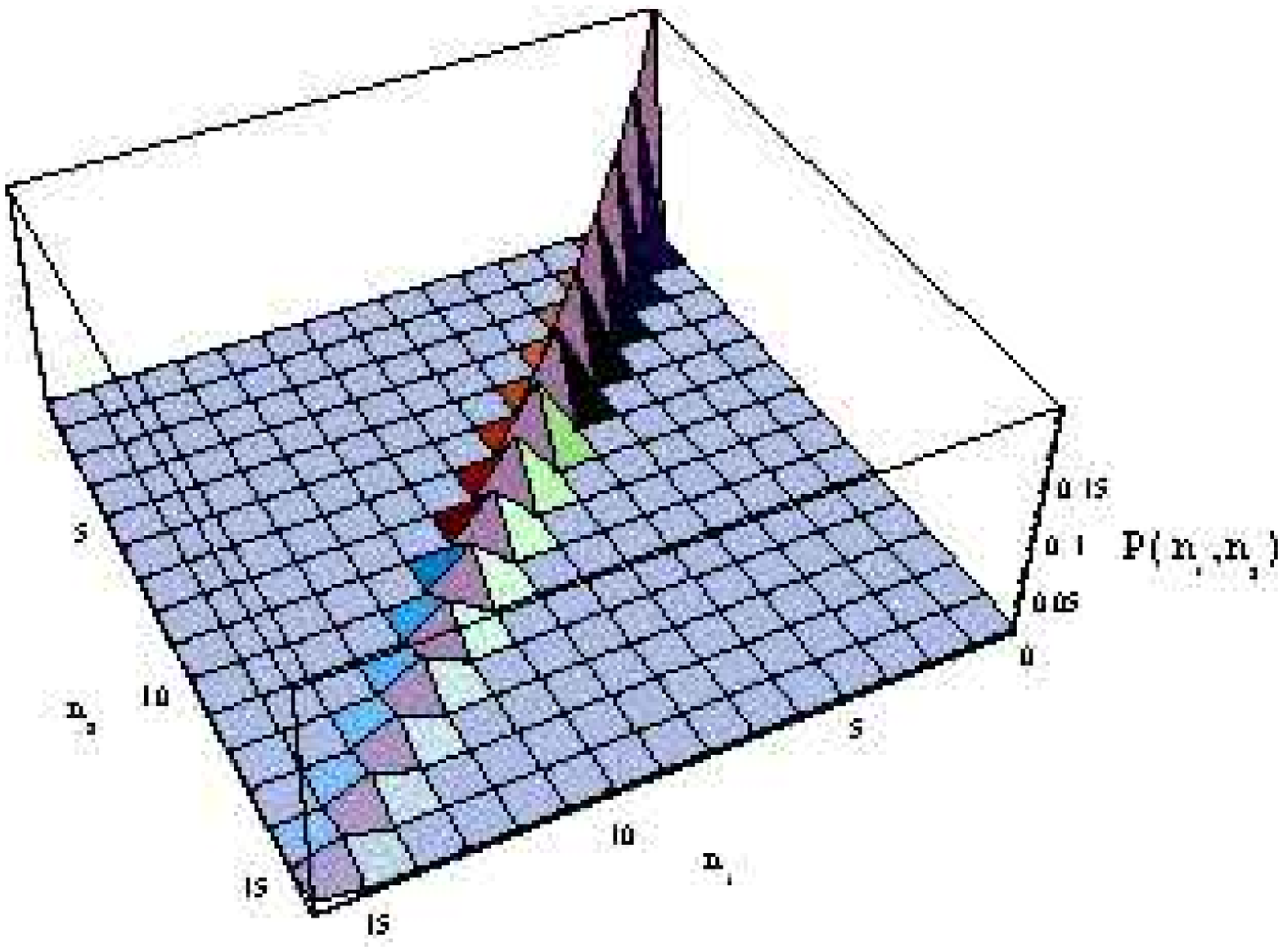}
\end{center}
\caption{$P(n_{1},n_{2})$ of the four-photon HEMPSS, for $r=1.5$,
$\beta_{1}=\beta_{2}=0$ and $|\gamma|=0.2$,
$\theta_{1}=\theta_{2}=0$, $\delta_{1}=\pi$, $\delta_{2}=0$.}
\label{pnhms11}
\end{figure}

As for the standard squeezed states, enhancing the squeezing
parameter $r$ enhances the oscillations of the PND. The effect of
the nonlinear strength $|\gamma |$ is competing with that of $r$.
In fact, we see that in Eq.~(\ref{nonlinstrength}) the effective
strength $\Delta$ of the nonlinear term involves the product
$|\gamma | e^{-r}$, with the choice of parameters yielding
$r>0$. Then, a sufficiently high value of $r$ reduces the effect
of increasing $|\gamma |$. If we look now at the symmetric PND's
(Figs.~\ref{pnhms1}, \ref{pnhms3}, \ref{pnhms5}, \ref{pnhms7}), we
see that, for a high value of $r$, increasing $|\gamma |$ from $0$
to $0.2$ does not change appreciably the oscillations (see
Figs.~\ref{pnhms3}, \ref{pnhms7} compared with Fig.~\ref{pnss2}).
On the contrary,
if $r$ is fixed to a lower value, increasing $|\gamma |$
results in a smoothening of the oscillations (Figs.~\ref{pnhms1},
\ref{pnhms5} compared with Fig. \ref{pnss1}). Obviously, when
$|\gamma |$ attains very high values, it cannot be contrasted by
$r$, and the oscillations are even more suppressed. In Figs.~\ref{pnhms10_2}
and \ref{pnhms11} we have plotted the PND for
$\beta_1 = \beta_2 = 0$, and for different values of $r$; this
corresponds to the two--mode four--photon squeezed vacuum. As in
the case of the standard two--mode squeezed vacuum \cite{PND}, we
obtain a diagonal, symmetric PND, in spite of the unbalanced
choice on the phases.

%%%%%%%%%%%%%%%%%%%%%%%%%%%%%%%%%%%%%%%%%%%%%%
%%%%%%%%%%%%%%%%%%%%%%%%%%%%%%%%%%%%%%%%%%%%%%

\begin{figure}
\begin{center}
\includegraphics*[width=8cm]{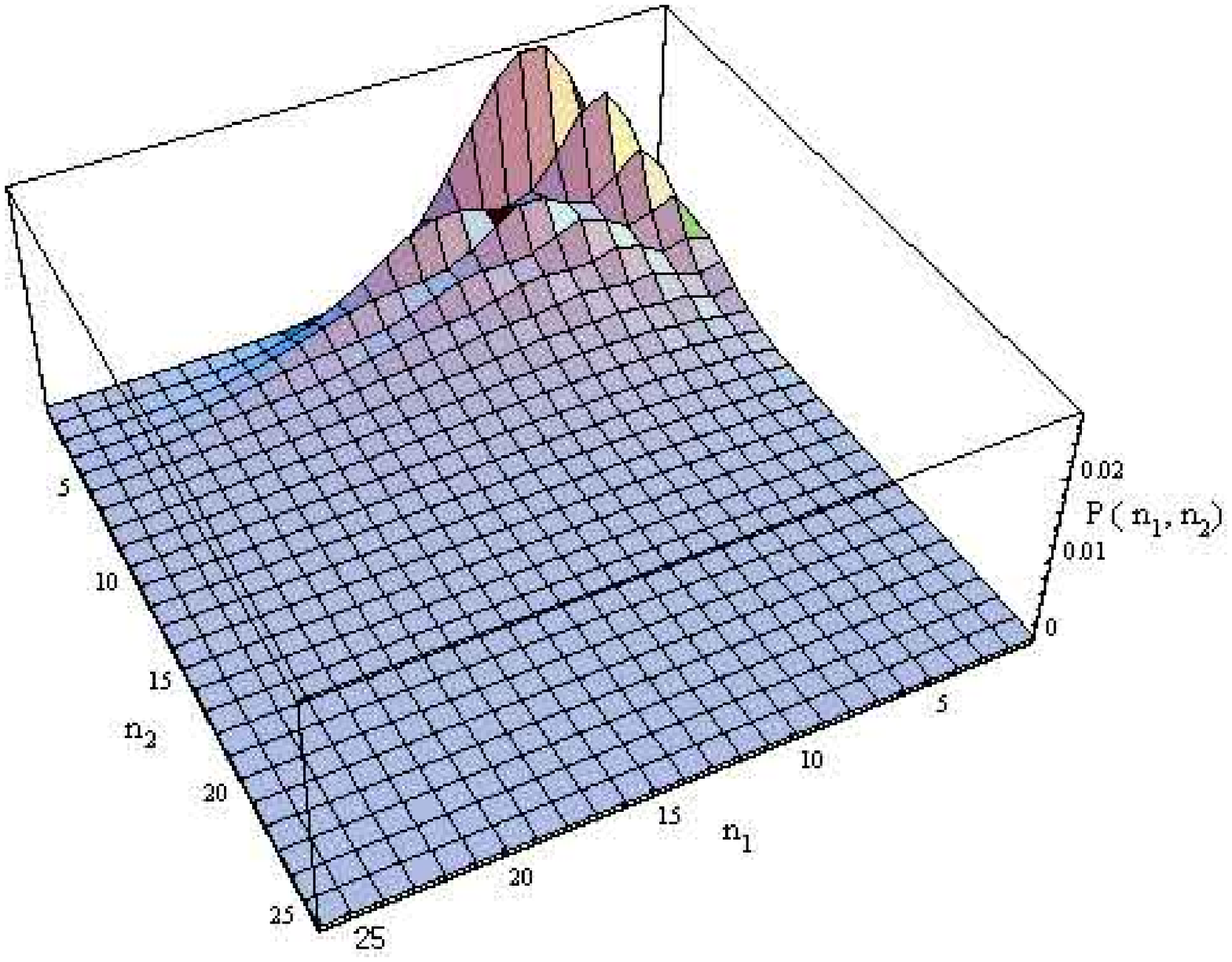}
\end{center}
\caption{$P(n_{1},n_{2})$ of the four-photon HEMPSS, for $r=0.8$,
$\beta_{1}=\beta_{2}=3$ and $|\gamma|=0.2$,
$\theta_{1}=-\theta_{2}=\frac{\pi}{6}$, $\delta_{1}=\pi$,
$\delta_{2}=0$.}
\label{pnhms12}
\end{figure}

In order to study the influence of the local oscillators angles,
in Fig.~\ref{pnhms12} we have plotted the PND for the same parameters
of unbalanced case of Fig.~\ref{pnhms6}, but with $\theta_1 = -
\theta_2 = \pi/6$; we see that the PND is modified with respect to
that of Fig.~\ref{pnhms6}, with a slight re--balancing between
the two axes, and less pronounced peaks.

%%%%%%%%%%%%%%%%%%%%%%%%%%%%%%%%%%%%%%%%%%%%%%%%%%%%%%%%%%%%%%

\begin{figure}
\begin{center}
\includegraphics*[width=8cm]{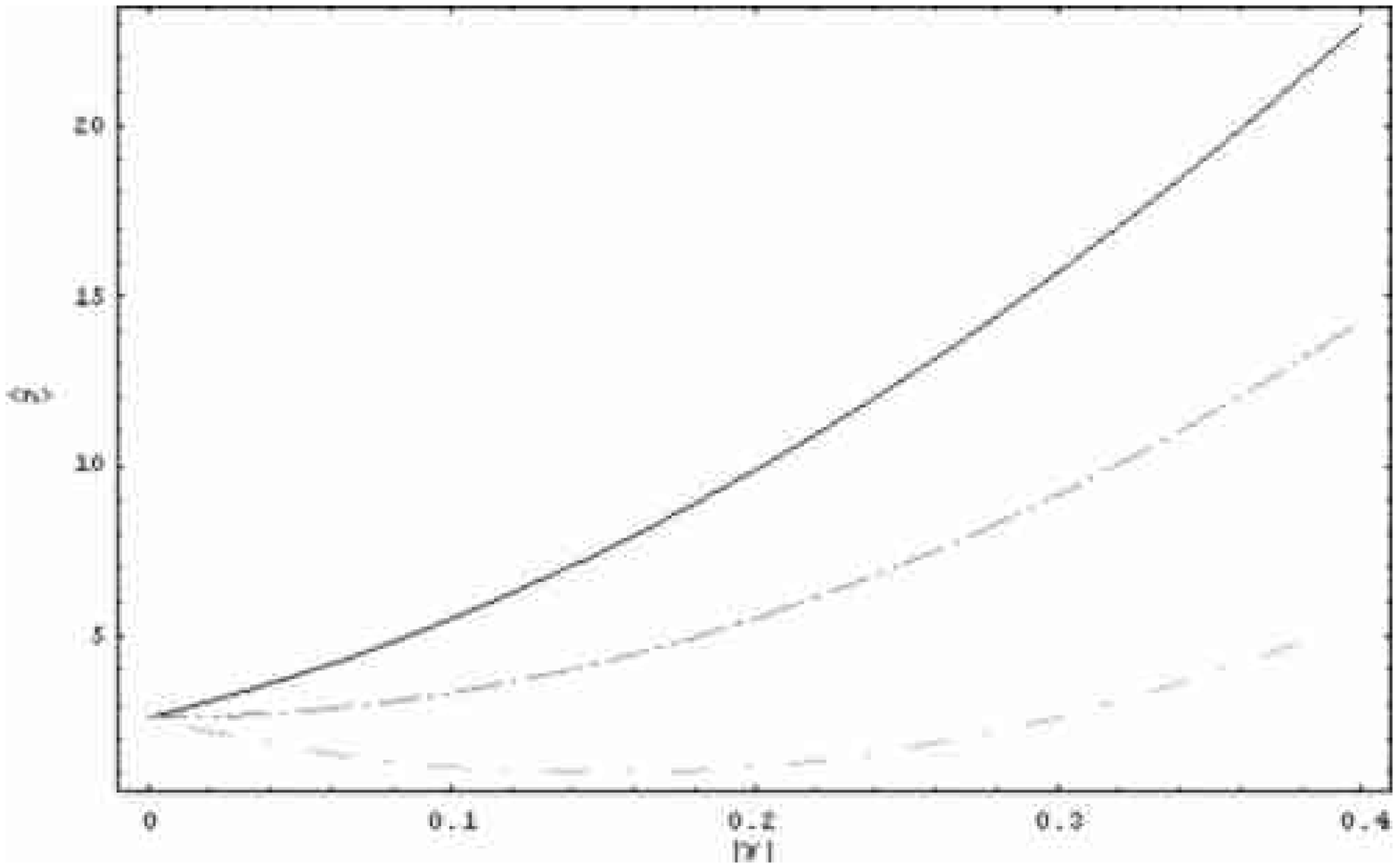}
\end{center}
\caption{$\langle n_{1}\rangle$ and $\langle n_{2}\rangle$ as
a function of $|\gamma|$, for the four-photon HEMPSS, at $r=0.8$,
$\beta_{1}=\beta_{2}=3$, $\theta_{1}=\theta_{2}=0$: 1)$\langle
n_{1}\rangle$ (full line) and $\langle n_{2}\rangle$ (dot--dashed
line) for $\delta_{1}=\pi$, $\delta_{2}=0$; 2)$\langle
n_{1}\rangle=\langle n_{2}\rangle$ for
$\delta_{1}=\delta_{2}=\frac{\pi}{2}$ (dotted line).}
\label{Nav1}
\end{figure}

\begin{figure}
\begin{center}
\includegraphics*[width=8cm]{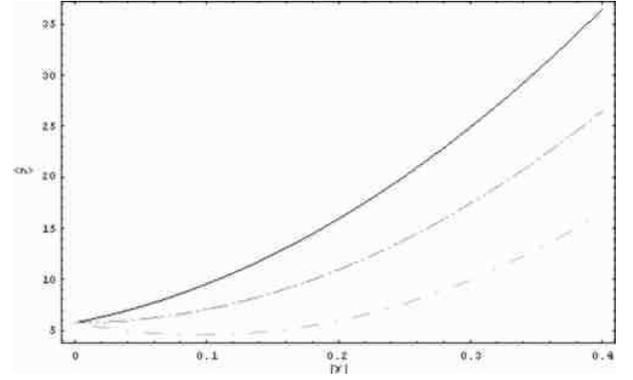}
\end{center}
\caption{$\langle n_{1}\rangle$ and $\langle n_{2}\rangle$ as
a function of $|\gamma|$, for the four-photon HEMPSS, at $r=1.5$,
$\beta_{1}=\beta_{2}=5$, $\theta_{1}=\theta_{2}=0$: 1)$\langle
n_{1}\rangle$ (full line) and $\langle n_{2}\rangle$ (dot--dashed
line) for $\delta_{1}=\pi$, $\delta_{2}=0$; 2)$\langle
n_{1}\rangle=\langle n_{2}\rangle$ for
$\delta_{1}=\delta_{2}=\frac{\pi}{2}$ (dotted line).}
\label{Nav2}
\end{figure}

In Figs.~\ref{Nav1} and \ref{Nav2} we show the mean
numbers of photons in the two modes as a function of $|\gamma |$ for
different values of the other parameters. We see that the balanced
choice on the phases gives $<n_1 > = <n_2 >$ (corresponding to
a symmetric PND), and the mean photon numbers increase monotonically
with increasing $|\gamma |$. In
the unbalanced instance (corresponding to a non symmetric PND), the
average photon number in the first mode is markedly larger and
monotonically increasing, while the average photon number in the
second mode first decreases, and increases monotonically beyond
a certain value of $|\gamma |$. This behavior is similar
to that exhibited by the HOMPSS for single-mode systems, as
discussed in Part I \cite{paper1}.

\begin{figure}
\begin{center}
\includegraphics*[width=8cm]{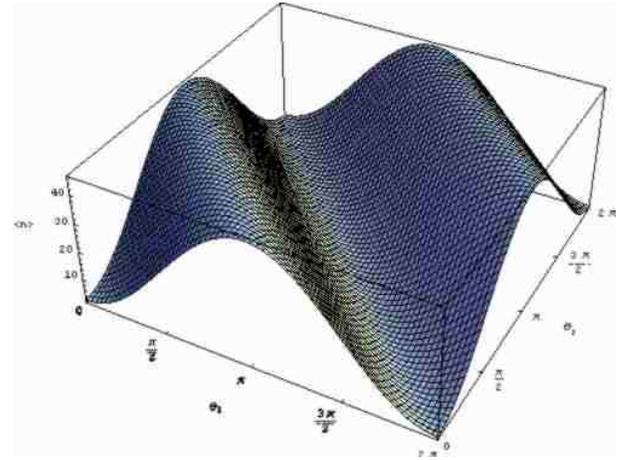}
\end{center}
\caption{$\langle n_{1}\rangle=\langle n_{2}\rangle$ as a function
of $\theta_1$ and $\theta_2$ ($\phi=\theta_1 +\theta_2$), of
a standard two-mode squeezed state ($\gamma=0$), for
$r=0.8$, $\beta_{1}=\beta_{2}=3$.}
\label{Navth}
\end{figure}

We next study the average photon number for the HEMPSS as
a function of the local oscillators angles $\theta_1$, $\theta_2$,
comparing with the case of two--mode squeezed states
(Fig.~\ref{Navth}). To this aim, when $|\gamma| = 0$ we retain as a
formal trick the dependence on $\theta_1$ and $\theta_2$ by fixing
$\phi=\theta_1 +\theta_2$, according to canonical conditions
Eqs.~(\ref{nlcc3}). Because of the choice $\beta_1
=\beta_2$, the plot $\langle n_{1}\rangle=\langle n_{2}\rangle$ is
somehow redundant; it is symmetric and shows an
oscillatory behavior, as it should be.

\begin{figure}
\begin{center}
\includegraphics*[width=8cm]{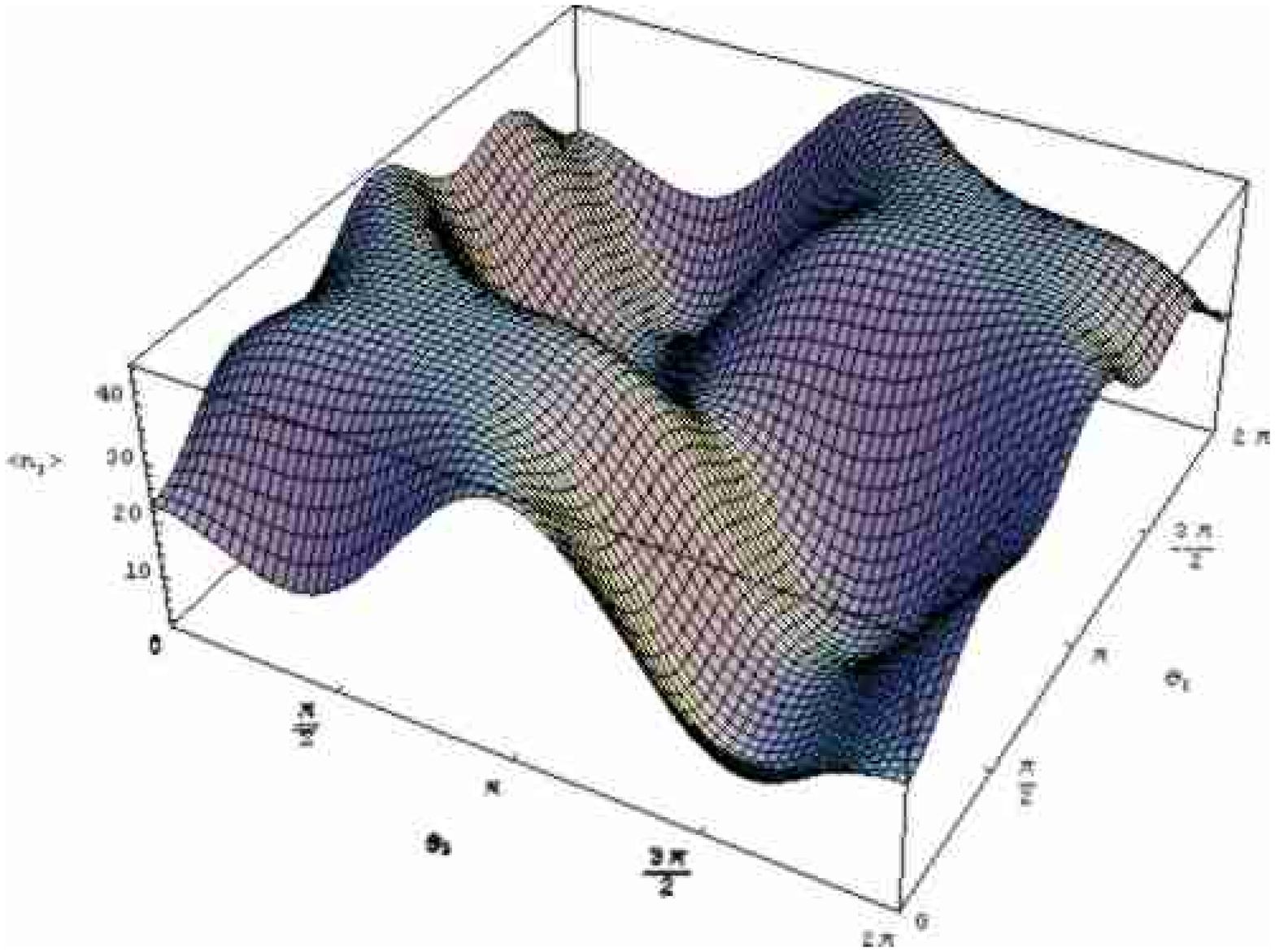}
\end{center}
\caption{$\langle n_{1}\rangle$ as a function of $\theta_1$ and
$\theta_2$, of the four-photon HEMPSS, for $r=0.8$, $\beta_{1}=\beta_{2}=3$,
$|\gamma|=0.4$, $\delta_1 =\pi$, $\delta_2 =\phi$.}
\label{Navth1}
\end{figure}

\begin{figure}
\begin{center}
\includegraphics*[width=8cm]{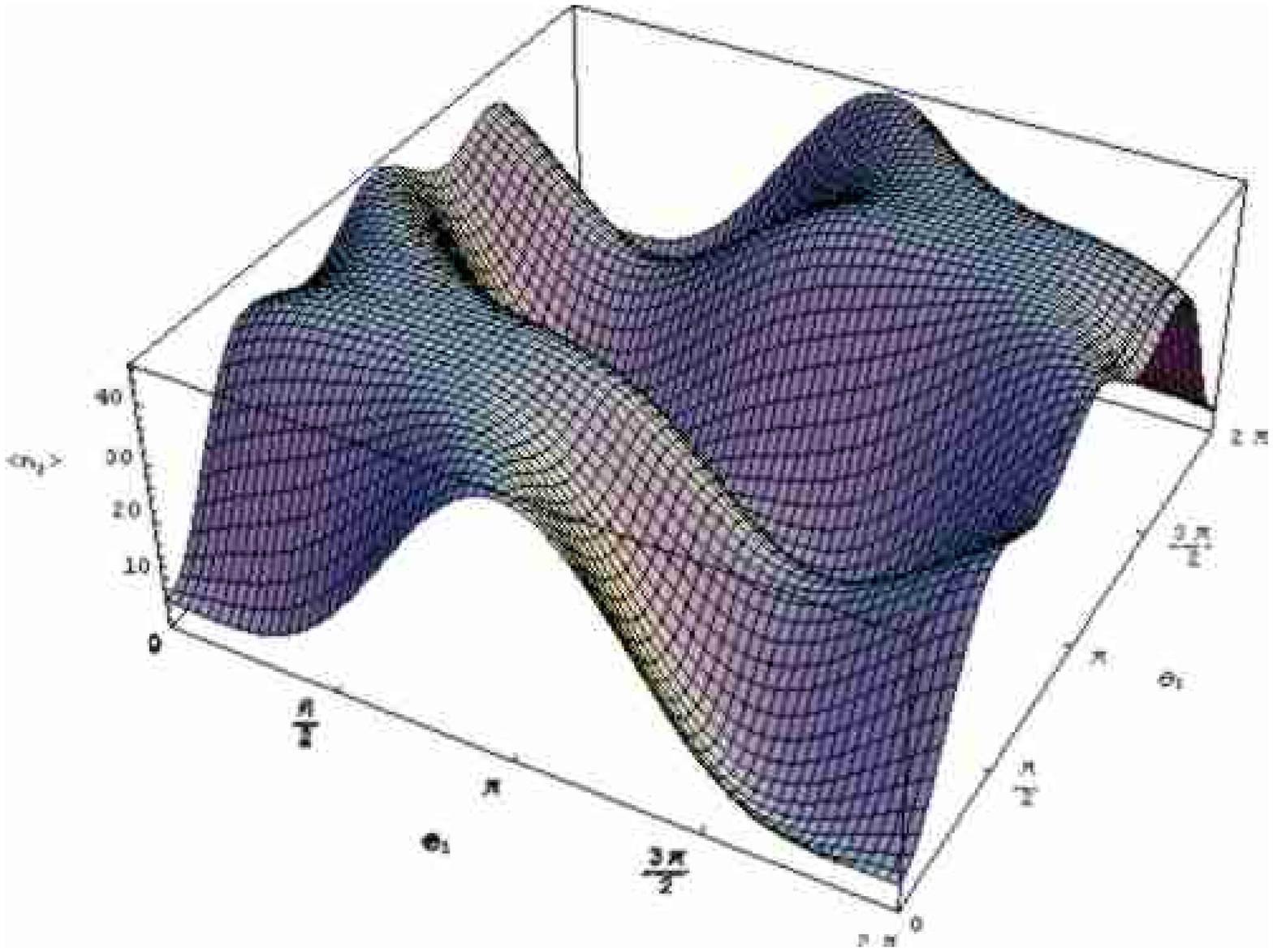}
\end{center}
\caption{$\langle n_{2}\rangle$ as a function of $\theta_1$ and
$\theta_2$, for the four-photon HEMPSS, at $r=0.8$, $\beta_{1}=\beta_{2}=3$,
$|\gamma|=0.4$, $\delta_1 =\pi$, $\delta_2 =\phi$.}
\label{Navth2}
\end{figure}

For the nonlinear case, we fix the parameters in the following
way: we let $\beta_{1}=\beta_{2}=3$, $r=0.8$ , $|\gamma|=0.4$ and
$\delta_1 =\pi$, $\delta_2 =\phi$, because of the canonical
conditions. As we can see in Figs.~\ref{Navth1} and \ref{Navth2},
the situation is very different from that of the standard two-mode
squeezed states. The shape is strongly
deformed and we can observe regions with a suppressed or enhanced
number of photons with respect to the mean number of a standard
two-mode squeezed state (see Fig.~\ref{Navth}).

\begin{figure}
\begin{center}
\includegraphics*[width=8cm]{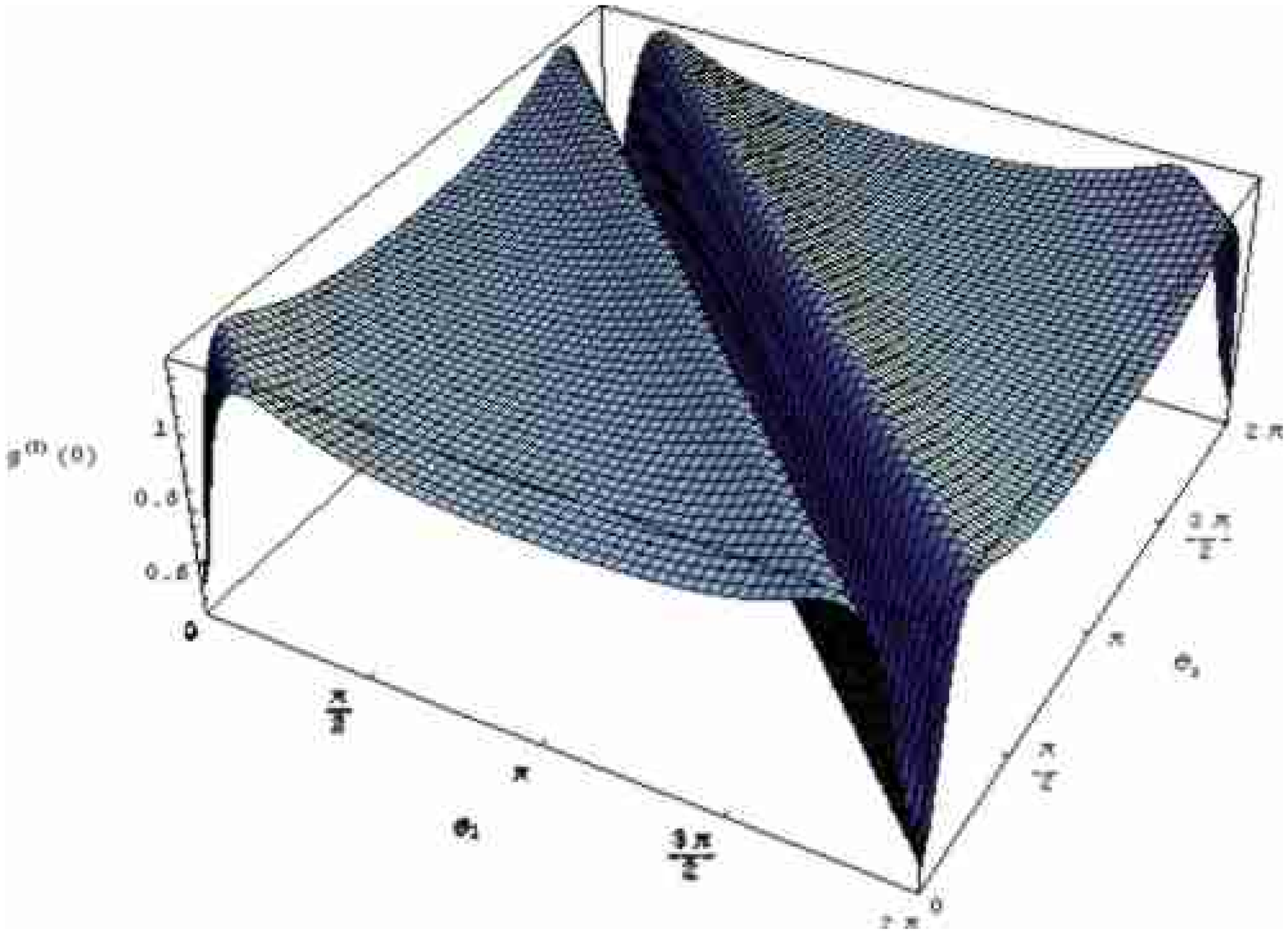}
\end{center}
\caption{$g^{(2)}(0)$ as a function of $\theta_{1}$ and
$\theta_{2}$($\phi=\theta_1 +\theta_2$), for a standard
two-mode squeezed state with $\gamma=0$, $r=0.8$, $\beta_{1}=\beta_{2}=3$.}
\label{g2_0}
\end{figure}

Finally, we have plotted the second-order
two-mode correlation function $g^{(2)}(0) =
\langle a_{1}^{\dagger}a_{1}a_{2}^{\dagger}a_{2} \rangle
/ \langle a_{1}^{\dagger}a_{1} \rangle
\langle a_{2}^{\dagger}a_{2} \rangle$ as a
function of the local
oscillators angles $\theta_1$ and $\theta_2$. We again obtain a
variety of shapes, corresponding to different statistical
behaviors. In particular we observe oscillating transitions
from bunching to antibunching behaviors. In this case as well,
we show for comparison the three-dimensional plot of $g^{(2)}(0)$
for a two--mode squeezed state Fig.~\ref{g2_0}. It is of course
symmetric and exhibits a narrow region of antibunching.

\begin{figure}
\begin{center}
\includegraphics*[width=8cm]{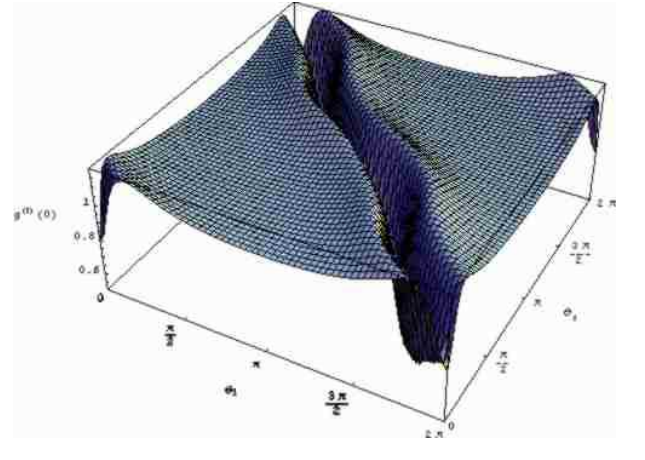}
\end{center}
\caption{$g^{(2)}(0)$ as a function of $\theta_{1}$ and
$\theta_{2}$, for the four-photon HEMPSS, at
$r=0.8$, $\beta_{1}=\beta_{2}=3$, $|\gamma|=0.1$,
$\delta_{1}=\pi$, and $\delta_{2}=\phi$.}
\label{g2_1}
\end{figure}

\begin{figure}
\begin{center}
\includegraphics*[width=8cm]{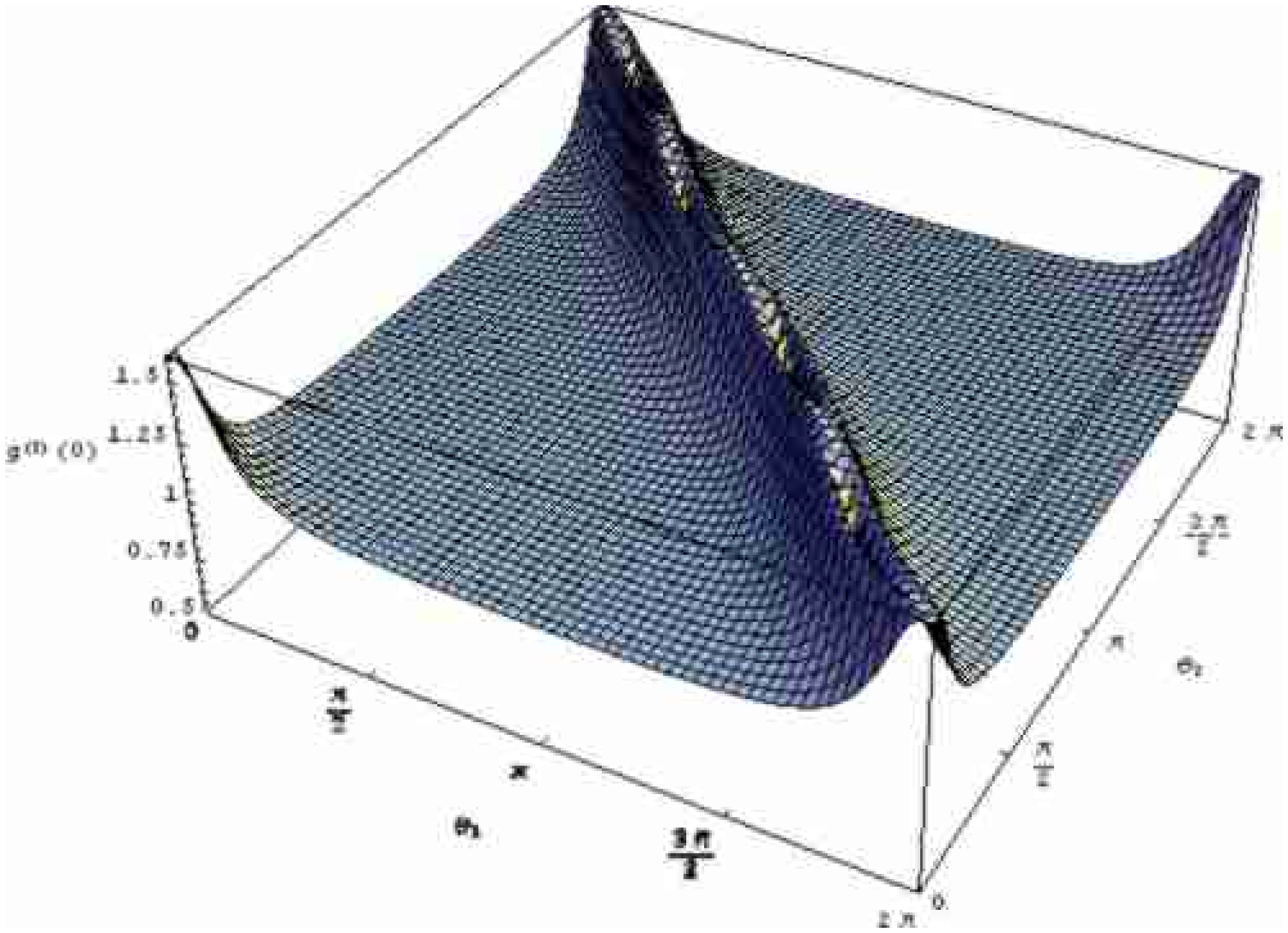}
\end{center}
\caption{$g^{(2)}(0)$ as a function of $\theta_{1}$ and
$\theta_{2}$, for the four-photon HEMPSS, at
$r=1.5$, $\beta_{1}=\beta_{2}=5$, $|\gamma|=0.1$,
$\delta_{1}=\pi$, and $\delta_{2}=\phi$.}
\label{g2_3}
\end{figure}

\begin{figure}
\begin{center}
\includegraphics*[width=8cm]{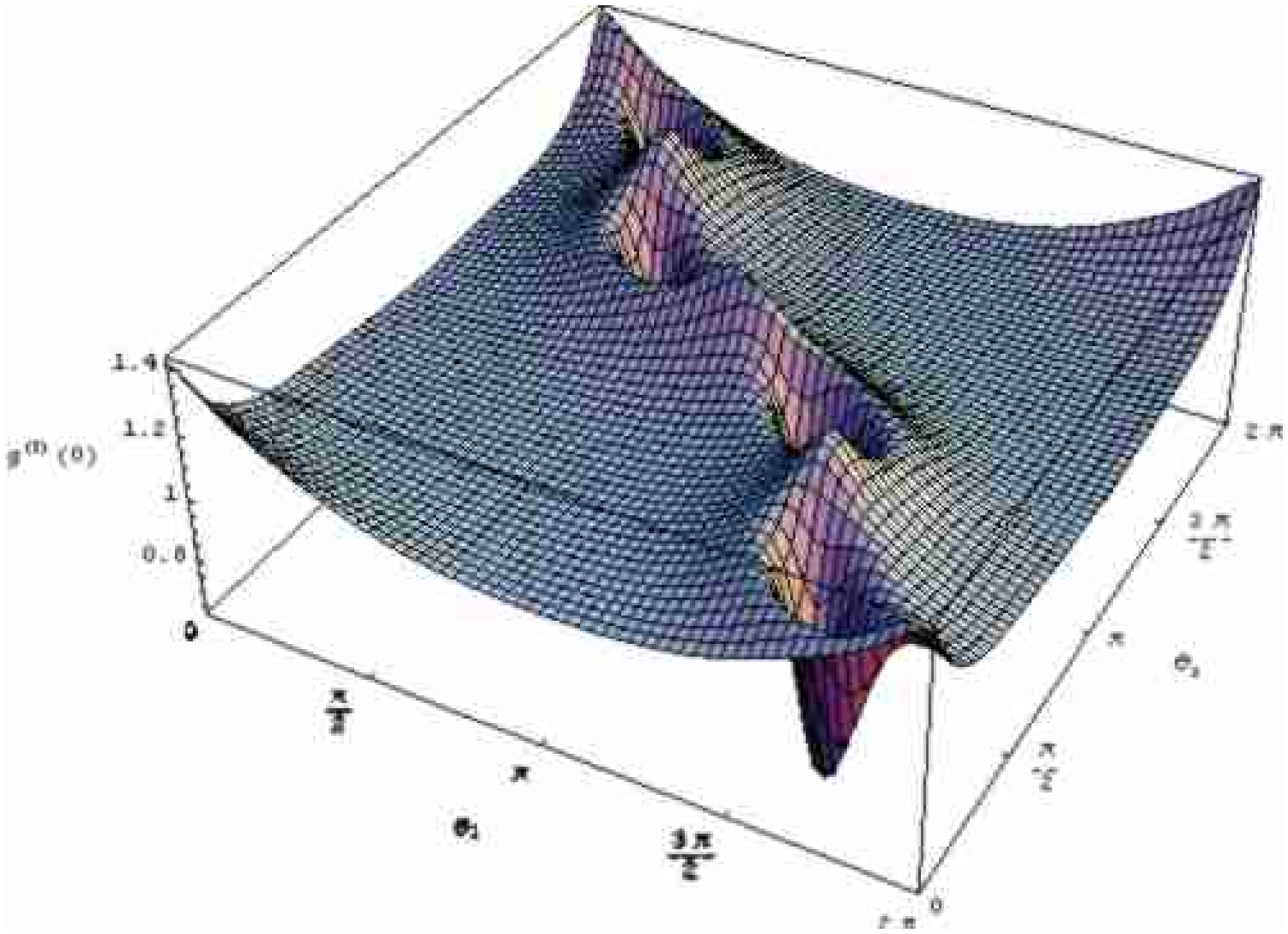}
\end{center}
\caption{$g^{(2)}(0)$ as a function of $\theta_{1}$ and
$\theta_{2}$, for the four-photon HEMPSS, at
$r=0.8$, $\beta_{1}=\beta_{2}=3$, $|\gamma|=0.2$,
$\delta_{1}=\pi$, and $\delta_{2}=\phi$.}
\label{g2_2}
\end{figure}

\begin{figure}
\begin{center}
\includegraphics*[width=8cm]{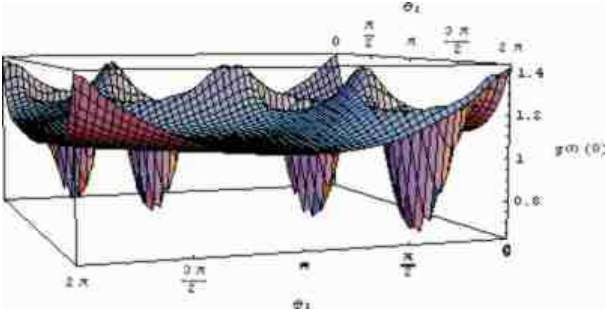}
\end{center}
\caption{$g^{(2)}(0)$ as a function of $\theta_{1}$ and
$\theta_{2}$, for the four-photon HEMPSS, with $r=0.8$,
$\beta_{1}=\beta_{2}=3$, $|\gamma|=0.2$,
$\delta_{1}=\pi$, and $\delta_{2}=\phi$.}
\label{g2_2bis}
\end{figure}

Already for small strengths of the nonlinearity ($|\gamma|=0.1$)
we can see from Fig.~\ref{g2_1} that the shape of the correlation
function is deformed with respect to that of the two-mode squeezed
state. This deformation becomes more marked for larger values
of $|\gamma|$, as shown in Figs.~\ref{g2_2} and \ref{g2_2bis}.
From Fig.~\ref{g2_3} we see that for
increasing values of the squeezing parameter, the antibunching
behavior is attenuated and the effect of the nonlinearity is
progressively suppressed. This is a further indication of the
competing effects of the squeezing and of the nonlinearity.

Summing up, similarly the case of single-mode systems, we observe that
the photon statistics of the two-mode, four-photon HEMPSS
can be strongly modified by the gauging of the parameters
entering the nonlinear canonical transformations, such as
the heterodyne mixing angles $\theta_1$ and $\theta_2$, the strength
$|\gamma|$, and the phases $\delta_1$ and $\delta_2$
of the nonlinearity.
Let us notice that this flexibility can be further
enhanced by considering the large number of different possible
canonical choices for the parameters that have not been
explored in the present work.

\section{Quantum dynamics in nonlinear media and generation
of the HEMPSS}

\label{nlqopt}

The interaction part of the Hamiltonian Eq.~(\ref{fpH}) contains
many non degenerate and degenerate multiphoton terms; this complex
structure is evidently due to the non trivial canonical
constraints associated to the structure of the nonlinear
Bogoliubov transformations. On the other hand, the very existence
of a canonical structure suggests that the experimental realization
of the associated multiphoton processes and squeezed states
should be conceivable. To this aim, we must bear in mind that
any experimental realization of the HEMPSS must necessarily
involve high order contributions to the susceptibility in nonlinear
media. In this Section we briefly review the essential
features of the theory of macroscopic quantum electrodynamics in
nonlinear media, referring for a more exhaustive treatment to
\cite{Shen,Tucker,Hillery}. The standard approach is the phenomenological
quantization of the classical macroscopic theory.
The macroscopic description of
the interaction of the electromagnetic field with matter is
based on the polarization vector \cite{Bloembergen1,Bloembergen2,Butcher}
\begin{eqnarray}
P_{i}&=&\underline{\chi}^{(1)}:\overline{E}
+\underline{\chi}^{(2)}:\overline{E}\overline{E}
+\underline{\chi}^{(3)}:\overline{E}\overline{E}\overline{E}+ ...
\;
\nonumber\\ && \nonumber\\
&=&\chi^{(1)}\delta_{ij}E_{j}+\chi^{(2)}_{ijk}E_{j}E_{k}
+\chi^{(3)}_{ijkl}E_{j}E_{k}E_{l} + ...
\end{eqnarray}
where $\overline{E}$ is the electric field, $\chi_{ijkl...}^{(n)}$
is the ($n+1)--th$ rank susceptibility tensor and the subscripts
indicate the spatial and polarization components. For a lossless,
nondispersive and uniform medium, the susceptibilities
$\underline{\chi}^{(n)}$ are symmetric tensors. Considering only
electric dipole interactions, the electric contribution to the
electromagnetic energy in the nonlinear medium is
\begin{equation}
H=\int_{V}d^{3}r\left[\frac{1}{2}\varepsilon_{0}E^{2}(\overline{r},t)+
\sum_{n}X_{n}(\overline{r})\right] \; ,
\label{Ham}
\end{equation}
where
\begin{equation}
X_{n}(\overline{r})=\frac{n\varepsilon_{0}}{n+1}\underline{\chi}^{(n)}:
\overline{E}\overline{E}...\overline{E} \; .
\end{equation}
In terms of the Fourier components of the field,
\bea
&&X_{n}=\frac{n\varepsilon_{0}}{n+1}\int \prod_{l = 1}^{n+1}
d \omega_{l}
e^{- i\omega_{l}t} \nonumber \\
&&\cdot \underline{\chi}^{(n)}(\omega_{2},...,\omega_{n+1}):
\overline{E}\overline{E}...\overline{E} \; . \label{Xn} \eea In
deriving Eq.~(\ref{Xn}), it is tacitly assumed that
$\chi^{n}(\omega)$ is independent on the wave-vector
$\overline{k}$ (electric dipole approximation), that in the
electromagnetic energy of the medium terms of order $\omega
\partial \chi^{(n)}/ \partial \omega$ can be ignored, and that
nonlinearities are small.

The canonical quantization of the macroscopic field in a nonlinear
medium is introduced by replacing the classical field
$\overline{E}(\overline{r},t)$ with the corresponding free-field
Hilbert space operator
\bea
\textbf{E}(\overline{r},t) &=&
i\sum_{\overline{k},\lambda}
\left[\frac{\hbar\omega_{\overline{k},\lambda}}
{2\varepsilon_{0}V}\right]^{1/2}
\{a_{\overline{k},\lambda}\epsilon_{\overline{k},\lambda}
e^{i(\overline{k}\cdot\overline{r}-\omega_{\overline{k},\lambda}t)}
\nonumber \\
\\
&&-a_{\overline{k},\lambda}^{\dag}\epsilon_{\overline{k},\lambda}^{*}
e^{-i(\overline{k}\cdot\overline{r}-\omega_{\overline{k},\lambda}t)}\}
\; , \nonumber
\eea
where $V$ is the spatial volume,
$\epsilon_{\overline{k},\lambda}$ is the unit vector encoding
the wave-vector
$\overline{k}$ and the polarization $\lambda$,
 $\omega_{\overline{k},\lambda}= c|\overline{k}|
/n_{\overline{k},\lambda}$
is the angular frequency in the medium ($n_{\overline{k},\lambda}$
being the index of refraction), and $a_{\overline{k},\lambda}$ is the
corresponding boson annihilation operator.
Denoting by $\bf{k}$ the pair
($\overline{k}, \lambda$), and defining
\begin{equation}
\Lambda_{\textbf{k}}=
\sqrt{\frac{\hbar\omega_{\bf{k}}}{2\varepsilon_{0}V}} \; ,
\end{equation}
the Fourier components of the quantum field are given by
\bea
\textbf{E}(\omega) &=& i\sum_{\bf{k}} \Lambda_{\textbf{k}}
\{a_{\bf{k}}\epsilon_{\bf{k}}
e^{i {\overline{k}}\cdot {\overline{r}}}\delta(\omega-\omega_{\bf{k}})
\nonumber \\
&& - a_{\bf{k}}^{\dag}\epsilon_{\bf{k}}^{*}
e^{-i {\overline{k}}\cdot {\overline{r}}}
\delta(\omega+\omega_{\bf{k}})\} \; .
\label{Four_comp}
\eea
The contribution of the $n$--th order nonlinearity to the
quantum Hamiltonian can thus be obtained by replacing the Fourier
components of the quantum field Eq.~(\ref{Four_comp})
in Eq.~(\ref{Xn}). Owing to the phase factors $e^{i\omega t}$, many
of the terms resulting from the Eq.~(\ref{Xn}) are rapidly
oscillating and average to zero; they can then be neglected
in the rotating wave approximation. The surviving terms correspond to
sets of frequencies that, due to the constraint of energy conservation,
satisfy the relation
\begin{equation}
\sum_{i=1}^{s}\omega_{\bf{k}_{i}}=\sum_{i=s+1}^{n+1}\omega_{\bf{k}_{i}}
\; ,
\label{freq_con}
\end{equation}
and involve products of boson operators of the form
\begin{equation}
a_{\bf{k}_{1}}a_{\bf{k}_{2}}\cdot\cdot\cdot a_{\bf{k}_{r}}
a_{\bf{k}_{r+1}}^{\dag}\cdot\cdot\cdot a_{\bf{k}_{n+1}}^{\dag} \; ,
\end{equation}
and their hermitian conjugates.
The occurrence of a particular multiphoton process is selected by imposing
the conservation of momentum. This is the so-called phase
matching condition and, classically, corresponds to the
synchronism of the phase velocity of the electric field and of the
polarization waves. These conditions can be realized by
exploiting the birefringent and dispersion properties of
anisotropic crystals.

The relevant modes of the radiation involved in a nonlinear
parametric process can be determined by the condition Eq.~(\ref{freq_con})
and the corresponding phase-matching condition
\be
\sum_{i=1}^{s}{{\overline{k}}}_{i}=
\sum_{i=s+1}^{n+1}{{\overline{k}}}_{i} \; .
\label{phas_match}
\ee
We will have to consider in detail the terms of the quantized
Hamiltonian Eq.~(\ref{Ham}) up to the fifth-order contribution
to the susceptibility tensor ($n=5$), using the expansion
Eq.~(\ref{Four_comp}) and assuming the matching conditions
Eqs.~(\ref{freq_con}) and (\ref{phas_match}).
These terms are needed to generate the processes
described by the Hamiltonian Eq.~(\ref{fpH}) and the
associated HEMPSS, and are listed in Appendix~\ref{Appendix}.

\section{Experimental setups}

\label{expsetup}

\subsection{Realizing the two--mode four--photon Hamiltonian}

\label{primasubsection}

We now specialize to a possible, but by no means unique,
experimental setup the previous
general results in the context of a multiple parametric
approximation, in order to realize the particular multiphoton
processes described by the two--mode Hamiltonians Eq.~(\ref{fpH}).

We consider a nonlinear crystal, illuminated by twelve laser pumps
at different frequencies :
\be
E^{(+)}=\sum_{i=1}^{12}E_{i}
e^{-i(\Omega_{i}t-\bf{k_{i}}\cdot \bf{r})} \, .
\ee
We assume a lossless, nondispersive, uniform, noncentrosymmetric
medium, in order to exploit the properties of spatial and Kleinman
symmetry of the susceptibility tensors, which can be taken to
be real.
Due to the high intensity of the lasers,
all the pump fields
are treated classically in this model. For simplicity, we will
assume a single polarization of the light,
and unidirectional propagating waves. Two
quantum modes of the radiation field, respectively at incommensurate
frequencies $\omega_{1}$ and $\omega_{2}$, are
excited in the parametric processes in the nonlinear crystal.
We impose the following relations on the
frequencies of the laser pumps :
\bea
&&\Omega_{1}+\Omega_{2}=\omega_{1}+\omega_{2} \; , \;
\quad \Omega_{3}+\Omega_{4}=2(\omega_{1}+\omega_{2}) \, ,
\nonumber \\
&& \nonumber \\ &&\Omega_{5}+\Omega_{6}=3\omega_{1} \; , \;
\quad \Omega_{7}+\Omega_{8}=3\omega_{2} \, ,
\nonumber \\
&&\Omega_{9}+\Omega_{10}=2\omega_{1}-\omega_{2} \, , \; \;
\Omega_{11}+\Omega_{12}=2\omega_{2}-\omega_{1} \, ,
\label{freqcond}
\eea
with the corresponding phase-matching conditions. With this choice,
given $\omega_{1}$ and $\omega_{2}$, each pair
of frequencies is fixed by the corresponding phase matching
condition. With a sufficiently large
number of laser pumps, many related
independent equations can be satisfied.
Reducing the number of pumps and of
independent equations is certainly possible, at the price
of leading to more complicated relations between the frequencies.

The introduction of several pumps, with suitable frequencies, allows to study
the multiphoton processes corresponding to higher-order contributions
to the susceptibility tensor.  We will use conditions
Eqs.~(\ref{freqcond}) to obtain two-, three- and four- photon down
conversion processes, and the associated mixed processes in the
interacting Hamiltonian. We write all the contributions considered
in our model in the interaction representation. Let us denote
by $\kappa^{js}_{lm}$ the coefficient of the generic interaction term
$a_{1}^{\dag j}a_{2}^{\dag s}a_{1}^{l}a_{2}^{m}$, describing the
creation of $j$ photons and the annihilation of $l$ photons in the
first mode, and the creation of $s$ photons and the annihilation
of $m$ photons in the second mode. The two--photon down
conversion term due to third-order interaction reads:
\be
H_{2}=\kappa^{11}_{00}a_{1}^{\dag}a_{2}^{\dag}E_{1}E_{2} + h.c. \, ,
\label{dc2}
\ee
with
$$
\kappa^{11}_{00}\propto\chi^{(3)}(-\Omega_{2};
-\omega_{1},-\omega_{2},\Omega_{1}).
$$
The degenerate three--photon down conversion terms are
\be
H_{3d}=(\kappa^{30}_{00}a_{1}^{\dag
3}E_{5}E_{6}+h.c.)+(\kappa^{03}_{00}a_{2}^{\dag 3}E_{7}E_{8}+h.c.)
\, ,
\label{dc3}
\ee
with
$$
\kappa^{30}_{00}\propto\chi^{(4)}(-\Omega_{6};-\omega_{1},
-\omega_{1},-\omega_{1},\Omega_{5}) \, ,
$$
and
$$
\kappa^{03}_{00}\propto\chi^{(4)}(-\Omega_{8};-\omega_{2},
-\omega_{2},-\omega_{2},\Omega_{7})\, .
$$
The semi-degenerate
four-photon down conversion term is given by:
\be
H_{4sd} =
\kappa^{22}_{00}a_{1}^{\dag 2}a_{2}^{\dag 2}E_{3}E_{4} + h. c.
\, ,
\label{dc4}
\ee
with
$$
\kappa^{22}_{00}\propto\chi^{(5)}(-\Omega_{4};-\omega_{1},-\omega_{1},
-\omega_{2},-\omega_{2},\Omega_{3})\, .
$$
The semi-degenerate three-photon terms are
\bea
&&H_{3sd}=\kappa^{21}_{00}a_{1}^{\dag 2}a_{2}E_{9}E_{10} + h. c. \, ,
\nonumber \\
&&H^{'}_{3sd}=\kappa^{12}_{00}a_{2}^{\dag 2}a_{1}E_{11}E_{12}+h.c. \, ,
\label{h12}
\eea
where
$$
\kappa^{21}_{00}\propto\chi^{(4)}(-\Omega_{10};-\omega_{1},
-\omega_{1},\omega_{2},\Omega_{9})\, ,
$$
and
$$
\kappa^{12}_{00}\propto\chi^{(4)}(-\Omega_{12};-\omega_{2},
-\omega_{2},\omega_{1},\Omega_{11})\; .
$$
In spite of their weaker contributions, we
retain for convenience the terms
\bea
H_{r} &&=
(\kappa^{21}_{10}a_{1}^{\dag 2}a_{2}^{\dag}a_{1}E_{1}E_{2} + h. c.)
\nonumber \\
&&+(\kappa^{12}_{01}a_{1}^{\dag}a_{2}^{\dag 2}a_{2}E_{1}E_{2} + h. c.)
\, ,
\label{H511}
\eea
with
$$
\kappa^{21}_{10}\propto
\chi^{(5)}(-\Omega_{2};
-\omega_{1},-\omega_{2},-\omega_{1},\omega_{1},\Omega_{1}) \, ,
$$
and
$$
\kappa^{12}_{01}\propto
\chi^{(5)}(-\Omega_{2};-\omega_{1},-\omega_{2},
-\omega_{2},\omega_{2},\Omega_{1})\, .
$$
Finally, we have to consider the Kerr contributions
\be
H_{kerr}=\kappa^{20}_{20}a_{1}^{\dag2}a_{1}^{2}
+\kappa^{02}_{02}a_{2}^{\dag2}a_{2}^{2}
+\kappa^{11}_{11}a_{1}^{\dag}a_{2}^{\dag}a_{1}a_{2} \, ,
\label{kerr}
\ee
where
$$
\kappa^{20}_{20}\propto\chi^{(3)}(-\omega_{1};-\omega_{1},
-\omega_{1},\omega_{1}) \, ,
$$
$$
\kappa^{02}_{02}\propto\chi^{(3)}(-\omega_{2};-\omega_{2},
-\omega_{2},\omega_{2}) \, ,
$$
and
$$
\kappa^{11}_{11}\propto\chi^{(3)}(-\omega_{2};-\omega_{1},
\omega_{1},-\omega_{2})\, .
$$
In conclusion, the full interaction Hamiltonian reads
\be
H_{I}=H_{2}+H_{3d}+H_{4sd}+H_{3sd}+H^{'}_{3sd}+H_{r}+H_{kerr} \,
\label{HI} \, .
\ee
We stress again that our choice is only one of the many
possible ones. One could envisage different
configurations, using nonlinear crystals of particular anisotropy
classes, and choosing different polarizations for the two
quantum modes.

Let us consider the interaction part of the Hamiltonian
Eq.~(\ref{fpH}). We see that all its terms are included in
Eq.~(\ref{HI}), and that the two Hamiltonians coincide
if we impose the equality
of the corresponding coefficients and assume
\be
\kappa^{20}_{20}\approx\kappa^{02}_{02}\approx 2\kappa^{11}_{11}
\, .
\ee
Equality of the coefficients can be realized by tuning
the complex amplitudes of the classical fields.

\vspace{5mm}

\subsection{Generating the four-photon HEMPSS}

\label{generating}

Eqs.~(\ref{unitoptot}) and (\ref{Uf2}) suggest a possible method to
realize experimentally the HEMPSS Eq.~(\ref{wf}) for
the simplest, quadratic nonlinearity.
Adopting the same criteria of Subsection~\ref{primasubsection}, we consider
again a sample nonlinear crystal, illuminated by eight classical
pumps at different frequencies
\be
E^{(+)}=\sum_{i=1}^{8}E_{i}
e^{-i(\Omega_{i}t-\bf{k_{i}}\cdot \bf{r})} \, .
\ee
We
impose the following relations
\bea
&&\Omega_{1}+\Omega_{2}=3\omega_{1} \, , \,
\quad \Omega_{3}+\Omega_{4}=3\omega_{2} \, , \nonumber \\
&&\Omega_{5}+\Omega_{6}=2\omega_{1}-\omega_{2} \, , \;
\; \; \Omega_{7}+\Omega_{8}=2\omega_{2}-\omega_{1} \, ,
\label{freqcond2}
\eea
and collinear phase-matching conditions.
We then obtain an interaction Hamiltonian of the form
\bea
&&H_{I}=\kappa^{30}_{00}a_{1}^{\dag3}E_{1}E_{2}+
\kappa^{03}_{00}a_{2}^{\dag3}E_{3}E_{4} \nonumber \\
&&+\kappa^{20}_{01}a_{1}^{\dag2}a_{2}E_{5}E_{6}
+\kappa^{02}_{10}a_{2}^{\dag 2}a_{1}E_{7}E_{8} + h. c. \, ,
\label{HI2}
\eea
with
$$
\kappa^{30}_{00}\propto\chi^{(4)}(-\Omega_{2};
-\omega_{1},-\omega_{1},-\omega_{1},\Omega_{1}) \, ,
$$

$$
\kappa^{03}_{00}\propto\chi^{(4)}(-\Omega_{4};
-\omega_{2},-\omega_{2},-\omega_{2},\Omega_{3})\, ,
$$

$$
\kappa^{20}_{01}\propto\chi^{(4)}(-\Omega_{6};
-\omega_{1},-\omega_{1},\omega_{2},\Omega_{5})\, ,
$$

$$
\kappa^{02}_{10}\propto\chi^{(4)}(-\Omega_{8};
-\omega_{2},-\omega_{2},\omega_{1},\Omega_{7}) \, ,
$$
where we have neglected the Kerr terms, due to their
weaker contributions.

Tuning the intensities of the external pumps, one can force the
Hamiltonian Eq.~(\ref{HI2}) to coincide with the interaction terms
appearing in the exponent of Eq.~(\ref{Uf2}).
The unitary operator Eq.~(\ref{Uf2})
can thus be realized by the Hamiltonian unitary evolution $e^{-i t
H_{I}}$, and the HEMPSS can be generated by imposing this unitary
Hamiltonian evolution on a standard two--mode squeezed state.

\section{Conclusions}

\label{sumout}

In this paper we define a canonical formalism for two--mode
multiphoton processes. We achieve this goal exploiting generalized,
nonlinear two--mode Bogoliubov transformations.
These transformations are defined by introducing a largely arbitrary,
nonlinear function of a heterodyne superposition of the
fundamental mode operators, depending on two local oscillator
angles (heterodyne mixing angles). The nonlinear transformations are
canonical once simple algebraic constraints, analogous to those
holding for the linear Bogoliubov transformations, are imposed
on the complex coefficients of the nonlinear mapping.
The scheme generalizes the one introduced in the companion
paper (Part I) \cite{paper1}. Among the
possible choices of the nonlinear function we pay special attention
to those associated to arbitrary $n-th$ powers of the heterodyne variables.
This class of transformations defines canonical
Hamiltonian models of $2n$--photons nondegenerate
and degenerate processes. We have determined the common
eigenstates of the transformed operators, and found their explicit
form in the entangled state representation. These states form
an overcomplete set and exhibit coherence
and squeezing properties. They are
non--Gaussian, entangled multiphoton states of
bipartite systems. They can be defined as well by
applying on the two-mode squeezed vacuum an unitary operator
with nonlinear exponent that realizes a heterodyne mixing of
the fundamental modes. We have thus named these states
Heterodyne Multiphoton Squeezed States (HEMPSS).
For quadratic nonlinearities, the HEMPSS realize a two--mode version
of the cubic phase states \cite{bartlett}.
Thus the HEMPSS are potentially
interesting candidates for schemes of quantum communication,
and for the implementation of continuous variable quantum computing
involving multiphoton processes.
One of the most appealing features of the HEMPSS concerns
the study of their statistical properties. As in the
single-mode case, but with even more striking effects, the
statistics can be largely gauged by tuning the
strength and the phases of the nonlinear interaction, and
the local oscillator (heterodyne mixing)
angles.
For the case of lowest (quadratic) nonlinearity,
we investigate possible routes to the experimental
realization of canonical multiphoton Hamiltonians, and the
generation of the associated HEMPSS.
Our proposal relies on the use of
higher-order contributions to the susceptibility in nonlinear media,
and on the engineering of suitable classical pumps and phase--matching
conditions.

In future work we intend to qualify and quantify the
entanglement of the HEMPSS, and their
possible use in the framework of the continuous
variables universal quantum computation. The
overcompleteness of the HEMPSS allows in principle to study the
unitary evolutions ruled by any diagonalizable
Hamiltonian associated to a quadratic nonlinearity, and by other
Hamiltonian operators associated to known group structures in
terms of the transformed mode variables.
In this way, we intend to introduce and characterize
other classes of nonclassical multiphoton states amenable
to analytic study.

\appendix

\section{}

\label{Appendix}

\noindent a) $X_{2}$ {\it contribution}

\vspace{3mm}

We fix
the frequencies $\omega_{\bf{k}_{1}}$, $\omega_{\bf{k}_{2}}$,
$\omega_{\bf{k}_{3}}$ such that
$\omega_{\bf{k}_{1}}=\omega_{\bf{k}_{2}}+\omega_{\bf{k}_{3}}$.
Ignoring the oscillating terms, we obtain
\begin{eqnarray}
&&\int_{V}d^{3}r X_{2}(\bf{r}) =
\frac{2}{3}\varepsilon_{0}\int_{V}d^{3}r\int (\prod_{l=1}^{3}
d\omega_{l}
e^{-i\omega_{l}t}) \nonumber \\
&&\cdot \underline{\chi}^{(2)}(\omega_{2},\omega_{3}):
\textbf{E}(\omega_{1})\textbf{E}(\omega_{2})\textbf{E}(\omega_{3})
\simeq \nonumber \\ && \nonumber
\\
&&2i\varepsilon_{0}
\sum_{\bf{k}_{1},\bf{k}_{2},\bf{k}_{3}}\Lambda_{\bf{k}_{1}}
\Lambda_{\bf{k}_{2}}\Lambda_{\bf{k}_{3}}
\underline{\chi}^{(2)}(\omega_{\bf{k}_{2}},\omega_{\bf{k}_{3}}):
\epsilon_{\bf{k}_{1}}^{*}\epsilon_{\bf{k}_{2}}\epsilon_{\bf{k}_{3}}
\nonumber \\
&&\cdot a_{\bf{k}_{1}}^{\dag}a_{\bf{k}_{2}}a_{\bf{k}_{3}}
\int_{V}d^{3}r
e^{-i(\bf{k}_{1}-\bf{k}_{2}-\bf{k}_{3})\cdot\bf{r}}+h.c.
\label{X2}
\end{eqnarray}
Condition Eq. (\ref{phas_match}) eliminates the strong dependence
on the phase mismatch
$\Delta\bf{k}=\bf{k}_{1}-\bf{k}_{2}-\bf{k}_{3}$ of the volume
integral Eq. (\ref{X2}).

The resulting nonlinear parametric processes (in a three wave
interaction) are described by a Hamiltonian of the form
\begin{equation}
H\propto\chi^{(2)}a^{\dag}bc+h.c. \; , \label{H2} \;
\end{equation}
where $a$, $b$, $c$ are three different modes at frequency
$\omega_{a}$, $\omega_{b}$, $\omega_{c}$, respectively.
Hamiltonian Eq. (\ref{H2}) can describe: sum-frequency mixing for
input $b$ and $c$ and $\omega_{a}=\omega_{b}+\omega_{c}$;
non-degenerate parametric amplification for input $a$, and
$\omega_{a}=\omega_{b}+\omega_{c}$; difference-frequency mixing
for input $a$ and $c$ and $\omega_{b}=\omega_{c}-\omega_{a}$.

If some of the modes in Hamiltonian Eq. (\ref{H2}) degenerate in
the same mode (i.e. at the same frequency, wave vector and
polarization), one obtains degenerate parametric processes as :
second harmonic generation for input $b=c$ and
$\omega_{a}=2\omega_{b}$; degenerate parametric amplification for
input $a$, and $\omega_{a}=2\omega_{b}$, with $b=c$; other effects
as optical rectification and Pockels effect involving d.c. fields.

\vspace{3mm}

\noindent b) $X_{3}$ {\it contribution}

\vspace{3mm}

We must now consider
\bea
&&\int_{V}d^{3}r X_{3}(\bf{r})=
\frac{3}{4}\varepsilon_{0}\int_{V}d^{3}r\int\ \prod_{l=1}^{4}
d\omega_{l}
e^{-i\omega_{l}t} \nonumber \\
&&\cdot \underline{\chi}^{(3)}(\omega_{2},\omega_{3},\omega_{4}):
\textbf{E}(\omega_{1})\textbf{E}(\omega_{2})\textbf{E}
(\omega_{3})\textbf{E}(\omega_{4}) \,.
\eea
In this case, the
relation Eq. (\ref{freq_con}) splits in the two distinct
conditions
\begin{eqnarray}
\omega_{\bf{k}_{1}}=\omega_{\bf{k}_{2}}
&+&\omega_{\bf{k}_{3}}+\omega_{\bf{k}_{4}}
\, , \label{Freq_con31} \;
\\ && \nonumber\\
\omega_{\bf{q}_{1}}+\omega_{\bf{q}_{2}}
&=&\omega_{\bf{q}_{3}}+\omega_{\bf{q}_{4}}
\; , \label{Freq_con32} \;
\end{eqnarray}
and the interaction Hamiltonian term takes the form \bea
&&H\propto[\alpha\chi^{(3)}(\omega_{\bf{k}_{2}},
\omega_{\bf{k}_{3}},\omega_{\bf{k}_{4}})
a_{\bf{k}_{1}}^{\dag}a_{\bf{k}_{2}}a_{\bf{k}_{3}}a_{\bf{k}_{4}}+h.c.]+
\nonumber \\ \nonumber
\\
&&[\beta\chi^{(3)}(-\omega_{\bf{q}_{2}},
\omega_{\bf{q}_{3}},\omega_{\bf{q}_{4}})
a_{\bf{q}_{1}}^{\dag}a_{\bf{q}_{2}}^{\dag}a_{\bf{q}_{3}}a_{\bf{q}_{4}}
+h.c.]\;, \label{H3} \; \eea where $\alpha$, $\beta$ are
$c$-numbers, and $a_{\bf{k}_{i}}$ ($i=1,..,4$) are four different
modes at frequency $\omega_{\bf{k}_{i}}$.

This term can give origin to a great variety of effects such as
high order harmonic generation, Kerr effect etc.

\vspace{5mm}

\noindent c) $X_{4}$ {\it contribution}

\vspace{5mm}

We manage now the term
\begin{eqnarray}
&&\int_{V}d^{3}r X_{4}(\bf{r})=
\frac{4}{5}\varepsilon_{0}\int_{V}d^{3}r\int \prod_{l=1}^{5}
d\omega_{l} e^{-i\omega_{1}t} \; \nonumber \\ \nonumber
\\
&&\cdot
\underline{\chi}^{(4)}(\omega_{2},\omega_{3},\omega_{4},\omega_{5}):
\textbf{E}(\omega_{1})\textbf{E}(\omega_{2})\textbf{E}
(\omega_{3})\textbf{E}(\omega_{4}) \textbf{E}(\omega_{5}) \;. \nonumber \\
\end{eqnarray}
Also in this case the relation (\ref{freq_con}) gives two
independent conditions:
\begin{eqnarray}
\omega_{\bf{k}_{1}}=\omega_{\bf{k}_{2}}
&+&\omega_{\bf{k}_{3}}+\omega_{\bf{k}_{4}}+
\omega_{\bf{k}_{5}} \; \label{Freq_con41} \, , \\
&& \nonumber\\
\omega_{\bf{q}_{1}}+\omega_{\bf{q}_{2}}
&=&\omega_{\bf{q}_{3}}+\omega_{\bf{q}_{4}}+
\omega_{\bf{q}_{5}} \; , \label{Freq_con42} \;
\end{eqnarray}
while the corresponding Hamiltonian contribution has a more
complex structure with multimode interactions:
\bea
&&H\propto[\delta\chi^{(4)}(\omega_{\bf{k}_{2}},\omega_{\bf{k}_{3}},
\omega_{\bf{k}_{4}},\omega_{\bf{k}_{5}})
a_{\bf{k}_{1}}^{\dag}a_{\bf{k}_{2}}a_{\bf{k}_{3}}a_{\bf{k}_{4}}a_{\bf{k}_{5}}
\nonumber \\ \nonumber
\\
&&+\gamma\chi^{(4)}(-\omega_{\bf{q}_{2}},
\omega_{\bf{q}_{3}},\omega_{\bf{q}_{4}},\omega_{\bf{q}_{5}})
a_{\bf{q}_{1}}^{\dag}a_{\bf{q}_{2}}^{\dag}
a_{\bf{q}_{3}}a_{\bf{q}_{4}}a_{\bf{q}_{5}}+h.c.] \, . \nonumber \\
\label{H4}
\eea
Here $\delta$, $\gamma$ are $c$-numbers, and
$a_{\bf{k}_{i}}$ ($i=1,..5$) are five different modes at frequency
$\omega_{\bf{k}_{i}}$.

\vspace{5mm}

\noindent d) $X_{5}$ {\it contribution}

\vspace{5mm}

At last we consider the term
\begin{eqnarray}
&&\int_{V}d^{3}r X_{5}(\bf{r})= \nonumber \\
&&\frac{5}{6}\varepsilon_{0}\int_{V}d^{3}r\int \prod_{l=1}^{6}
d\omega_{l} e^{-i\omega_{1}t}
\underline{\chi}^{(5)}(\omega_{2},\omega_{3},\omega_{4},\omega_{5},\omega_{6}):
\; \nonumber \\ \nonumber
\\
&& \textbf{E}(\omega_{1})\textbf{E}(\omega_{2})\textbf{E}
(\omega_{3})\textbf{E}(\omega_{4})
\textbf{E}(\omega_{5})\textbf{E}(\omega_{6}) \;.
\end{eqnarray}
In this case the relation (\ref{freq_con}) gives three conditions:
\begin{eqnarray}
\omega_{\bf{k}_{1}}=\omega_{\bf{k}_{2}}
&+&\omega_{\bf{k}_{3}}+\omega_{\bf{k}_{4}}+ \omega_{\bf{k}_{5}}+
\omega_{\bf{k}_{6}}
\label{Freq_con51} \, , \\
&& \nonumber\\
\omega_{\bf{q}_{1}}+\omega_{\bf{q}_{2}}
&=&\omega_{\bf{q}_{3}}+\omega_{\bf{q}_{4}}+ \omega_{\bf{q}_{5}}+
\omega_{\bf{q}_{6}}
\label{Freq_con52} \, , \\
&& \nonumber\\
\omega_{\bf{p}_{1}}+\omega_{\bf{p}_{2}}
&+&\omega_{\bf{p}_{3}}=\omega_{\bf{p}_{4}}+ \omega_{\bf{p}_{5}}+
\omega_{\bf{p}_{6}} \label{Freq_con53} \; .
\end{eqnarray} The corresponding Hamiltonian contribution is:
\bea &&H\propto[\eta_1
\chi^{(5)}(\omega_{\bf{k}_{2}},\omega_{\bf{k}_{3}},
\omega_{\bf{k}_{4}},\omega_{\bf{k}_{5}},\omega_{\bf{k}_{6}})
a_{\bf{k}_{1}}^{\dag}a_{\bf{k}_{2}}a_{\bf{k}_{3}}a_{\bf{k}_{4}}a_{\bf{k}_{5}}a_{\bf{k}_{6}}
\nonumber \\ && \nonumber
\\
&&+\eta_2 \chi^{(5)}(-\omega_{\bf{q}_{2}},
\omega_{\bf{q}_{3}},\omega_{\bf{q}_{4}},\omega_{\bf{q}_{5}},\omega_{\bf{q}_{6}})
a_{\bf{q}_{1}}^{\dag}a_{\bf{q}_{2}}^{\dag}
a_{\bf{q}_{3}}a_{\bf{q}_{4}}a_{\bf{q}_{5}}a_{\bf{q}_{6}}
\nonumber \\ && \nonumber \\
&&+\eta_3 \chi^{(5)}(-\omega_{\bf{p}_{2}},-\omega_{\bf{p}_{3}},
\omega_{\bf{p}_{4}},\omega_{\bf{p}_{5}},\omega_{\bf{p}_{6}})
a_{\bf{p}_{1}}^{\dag}a_{\bf{p}_{2}}^{\dag}a_{\bf{p}_{3}}^{\dag}
a_{\bf{p}_{4}}a_{\bf{p}_{5}}a_{\bf{p}_{6}} \nonumber \\&&
\nonumber \\
&&+h.c.] \label{H5} \eea

Here $\eta_i$ ($i=1,2,3$) are $c$-numbers, and $a_{\bf{k}_{i}}$
($i=1,..6$) are six different modes at frequency
$\omega_{\bf{k}_{i}}$.

\end{document}